\documentclass[11pt,preprint]{aastex}

\shorttitle{Radio-loudness of AGNs}
\shortauthors{Sikora, Stawarz, \& Lasota}

\newcommand {\be} {\begin{equation}}
\newcommand {\ee} {\end{equation}}

\begin{document}

\title{Radio-Loudness of Active Galactic Nuclei:\\
Observational Facts and Theoretical Implications}

\author{Marek Sikora$^{1,2,3}$, {\L}ukasz Stawarz$^{3,4,5}$, and Jean-Pierre Lasota$^{2,4}$}
\affil{
$^1$Nicolaus Copernicus Astronomical Center, Bartycka 18, 00-716
Warsaw, Poland\\
$^2$Institut d'Astrophysique de Paris, UMR 7095 CNRS, Universit\'e
Pierre et Marie Curie, 98bis Bd Arago, 75014 Paris, France\\
$^3$Kavli Institute for Particle Astrophysics and Cosmology, Stanford University,
Stanford CA 94305, and Stanford Linear Accelerator Center, MenloPark CA 94025\\
$^4$Astronomical Observatory, Jagiellonian University, ul. Orla 171, 30-244 Krak\'ow, Poland\\
$^5$Landessternwarte Heidelberg, K\"onigstuhl, and Max-Planck-Institut f\"ur Kernphysik, Saupfercheckweg 1, Heidelberg 69117, Germany}
\email{sikora@camk.edu.pl , stawarz@slac.stanford.edu , lasota@iap.fr}

\begin{abstract}
We investigate how the total radio luminosity of AGN-powered radio
sources depends on their accretion luminosity and the central black
hole mass. Our studies cover about seven orders of magnitude in
accretion luminosity (expressed in Eddington units, i.e. as
Eddington ratios) and the full range of AGN black hole masses. We
find that AGNs form two distinct and well separated sequences on the
radio-loudness -- Eddington-ratio plane. The `upper' sequence is
formed by radio selected AGNs, the `lower' sequence contains mainly
optically selected objects. Whereas an apparent `gap' between the
two sequences may be an artifact of selection effects, the sequences
themselves mark the real upper bounds of radio-loudness of two
distinct populations of AGNs: those hosted respectively by
elliptical and disk galaxies. Both sequences show the same
dependence of the radio-loudness on the Eddington ratio (an increase
with decreasing Eddington ratio), which suggests that the
normalization of this dependence is determined by the black hole
spin. This implies that central black holes in giant elliptical
galaxies have (on average) much larger spins  than black holes in
spiral/disc galaxies. This galaxy-morphology related radio-dichotomy
breaks down at high accretion rates where the dominant fraction of
luminous quasars hosted by elliptical galaxies is radio quiet. This
led to speculations in the literature that formation of powerful
jets at high accretion rates is intermittent and related to switches
between two disk accretion modes, as directly observed in some BH
X-ray binaries. We argue that such intermittency can be reconciled
with the spin paradigm, provided that successful formation of
relativistic jets by rotating black holes requires collimation by
MHD outflows from accretion disks.
\end{abstract}


\section{Introduction}

It took less than two years from the discovery of the first quasars
to realize that most of them are radio-quiet rather than radio-loud
(Sandage 1965). Strittmatter et al. (1980) pointed out that
radio-loudness of quasars, defined as the radio-to-optical flux
density ratio, may have a bimodal distribution. Radio bimodality was
confirmed by Kellermann et al. (1989), who demonstrated that there
are at least $5-10$ times more radio-quiet than radio-loud quasars
(see also Miller, Peacock, \& Mead 1990; Stocke et al. 1992).
However, most recent studies based on deep radio surveys FIRST and
NVSS (Becker, White, \& Helfand 1995; Condon et al. 1998), and
optical massive surveys SDSS and 2dF  (York et al. 2000; Croom et
al. 2001), confirm very broad distribution of radio-loudness among 
quasars but are not very conclusive about bimodality nature of this 
distribution (White et al. 2000; Ivezi\'c et al. 2002; Cirasuolo et 
al. 2003a,b; Laor 2003).

Recently, the issue of radio-loudness got a new dimension: after
astronomers had learned how to `weigh' supermassive black holes (see
Woo \& Urry 2002 and refs. therein), it became possible to study the
dependence of the radio-loudness parameter, ${\cal R} \equiv
L_{\nu_R}/L_{\nu_{opt}}$, on the Eddington ratio, $\lambda \equiv
L_{bol}/L_{Edd}$. Here $L_{\nu_R}$ and $L_{\nu_{opt}}$ stand for the
monochromatic luminosities at some specified radio, $\nu_R$, and
optical, $\nu_{opt}$, frequencies, while $L_{bol}$ and $L_{Edd}$
denote the bolometric luminosity of the active nucleus and the
appropriate Eddington luminosity ($L_{\rm Edd}=4\pi G{\cal M} m_{p}
c/\sigma_T = 1.3 \times 10^{38} {\cal M}/{\cal M_{\odot}} \ \rm erg\
s^{-1}$), respectively. The analysis of radio-loudness for PG
quasars, Seyfert galaxies, and LINERs, performed by  Ho (2002) seem
to indicate that radio-loudness increases with decreasing Eddington
ratio, however, with a huge scatter in ${\cal R}$. His results
showed also that the largest ${\cal R}$ are found in AGNs with black
hole masses $\gtrsim 10^8 {\cal M}_{\odot}$. Such a mass-related
duality was recently confirmed by Chiaberge, Capetti, \& Macchetto
(2005) who included FR I radio-galaxies in their sample. The effect,
however, could not be clearly identified for intermediate Eddington
ratios. We will show that the reason for the above is that those
studies did not take into account luminosities of extended radio
structures (Laor 2004) and did not include broad-line radio galaxies
(hereafter BLRGs). These objects are almost exclusively hosted by
giant elliptical galaxies and are about $10^3$ times radio louder
then Seyfert galaxies for the same range of the Eddington ratio. The
criteria for selection of BLRGs used in our studies are specified in
\S2. These sources are studied together with Seyfert galaxies,
radio-quiet LINERs, FR I radio galaxies, and quasars. The results
are presented in \S3.

The studies of radio-loudness are crucial for addressing such basic
questions as how jets are  formed, accelerated and collimated, and
why the efficiency of  jet production can be so different
 among objects  very similar in all other aspects. The
same questions concern jets in black-hole and neutron-star X-ray
binaries (XRBs; see review by Fender 2004). Taking advantage of the
very short time-scales of XRB variabilities, the dependence of the
radio-loudness on the Eddington ratio in these objects can be traced
directly, for each source individually. Such studies  indicate that
at low luminosities the radio-loudness is a monotonic function of
the accretion luminosity (Gallo, Fender \& Pooley 2003), while at
the highest luminosities it may jump by a large factor, following
transitions between two accretion states (Fender, Belloni, \& Gallo
2004). If the radio activity of individual AGNs depends on the
Eddington ratio in a similar way, then the observed huge differences
of radio-loudness for AGNs with similar Eddington ratio, especially
at its lowest values, indicates that yet another parameter in
addition to the accretion rate must play a role in determining the
jet production efficiency. In \S4 we investigate the possibility
that this parameter is the black-hole spin and speculate how
the spin paradigm can be reconciled with intermittent jet activity
at higher accretion rates.  Our main results and their theoretical
implications are summarized in \S5.

In this paper we assume $\Lambda$CDM cosmology, with $\Omega_M=0.3$,
$\Omega_{\Lambda}=0.7$, and the Hubble constant $H_0=70$ km s$^{-1}$
Mpc$^{-1}$.

\section{Samples}
Our studies include: radio-loud broad-line AGNs (BLRGs plus
radio-loud quasars);  Seyfert galaxies and LINERs; FR I radio
galaxies; optically selected quasars. The subsamples were selected
according to the following criteria:
\begin{itemize}
\item the optical flux of the central, unresolved source is known;
\item the total radio flux is known (including extended emission if present);
\item black hole masses or necessary parameters to estimate them are available in literature.
\end{itemize}
\noindent Other criteria, applied individually to different
subsamples, are  specified below. In our sample, we did not include
blazars, i.e., OVV-quasars, HP-quasars, and BL Lac objects, because
their observed emission is significantly Doppler boosted. We also
did not analyze narrow line radio galaxies (NRLGs), because their
optical nuclei are hidden by ``dusty tori", which makes estimation
of the accretion rates very uncertain.

\subsection{Radio-selected broad line AGN}
The objects are taken from Eracleous \& Halpern (1994; 2003;
hereafter EH94 and EH03, respectively) who studied profiles of the
broad H$\alpha$ emission lines of radio-loud AGNs with $z \le 0.4$,
selected from V\'eron-Cetti \& V\'eron (1989). We divided the sample
in two sub-groups: the radio-loud quasars and BLRGs, with the
commonly used  division line at $M_V=-23$ which corresponds to the
$V$-band luminosity $L_V \simeq 10^{44.6} {\rm ergs \, s}^{-1}$.
They are listed in Tables 1 and 2, along with the following data:
IAU coordinates for the J2000.0 epoch; name of the source; redshift,
$z$, with the most accurate values taken from Eracleous \& Halpern (2004);
$V$-band total apparent magnitude, $m_V$, taken from EH94 and EH03; Galactic
extinction, $A_V$, available in NED
(http://nedwww.ipac.caltech.edu/); starlight contamination,
$\kappa_\star$, taken from EH94 and EH03; total radio flux at $\nu_5
\equiv 5$ GHz, $F_5$, obtained from the literature with references
provided in the tables; FWHM of the H$\alpha$ line taken from EH94
and EH03. We have used these data to calculate other quantities
included in the tables, namely: the optical luminosity of the
nucleus at $\lambda_B \equiv 4400$\AA, $L_B \equiv \nu_B L_{\nu_B}$;
the radio luminosity at $5$ GHz, $L_R \equiv \nu_5 L_{\nu_5}$; the
radio-loudness parameter, ${\cal R} \equiv L_{\nu_5}/L_{\nu_B} =
1.36 \times 10^5 (L_R/L_B)$; the black hole mass, ${\cal M}_{\rm
BH}$; and finally the $L_B$ and $L_R$ luminosities expressed in the
Eddington units.

The $B$-band nuclear luminosity was calculated using the standard
luminosity-flux relation:
\be L_B = 4 \pi d_L^2 \, \nu_B F_{\nu_B} \, (1+z)^{-(1-\alpha_{opt})} \quad , \ee
where $d_L$ is the luminosity distance calculated for a given
redshift  and the assumed cosmology, $\alpha_{opt}$ is the power-law
slope around $\nu_B$, and the $B$-band nuclear flux is
\be \nu_B F_{\nu_B} =
(\lambda_V/ \lambda_B)^{(1-\alpha_{opt})} \, [-0.4(m_V-A_V) -4.68] \,
(1-\kappa_{\star}) \quad , \ee
where $\lambda_V \equiv 5500$\AA, and $\alpha_{opt}=0.5$ is taken.
In this  paper we also assume the bolometric luminosity of the
active nucleus $L_{bol} = 10 \, L_B$. The total radio luminosity of
the source, $L_R$, was evaluated using a formula analogous to the
one given by Eq.(1), with the assumed radio spectral index $\alpha_R
= 0.8$ for the K-correction. We note, that most of the objects in
this subsample are strong radio sources ($F_5 > 0.03$ Jy), have
radio-morphologies of the FR II type and radio luminosities
dominated by the extended structures. Radio fluxes for most of them
were therefore taken from the single-dish radio surveys (Wright \&
Otrupcek 1990, Gregory \& Condon 1991). In the case of the three
weak sources IRAS 02366-3101, MS 0450.3-1817, and CBS 74, for which
the radio data from other facilities were used as indicated in the
tables, the provided radio fluxes (and hence $L_R$ and ${\cal R}$
parameters) should be rather considered as lower limits. Finally,
since black hole masses for most of the Eracleous and Halpern
objects are not available in literature, we estimated them using a
BLR size--luminosity relation assuming virial velocities of the gas
which produces broad H$\alpha$ lines (see, e.g., Woo \& Urry 2002
and references therein),
\be {{\cal M}_{\rm BH} \over {\cal M_{\odot}}} = 4.8 \times
\left[\lambda L_{\lambda} (5100{\rm \AA}) \over 10^{44}{\rm
ergs/s}\right]^{0.7} \, {\rm FWHM}_{{\rm H}\alpha}^2 \quad , \ee
where FWHM$_{{\rm H}\alpha}$ is derived by EH94 or EH03, and $
L_{\lambda}(5100{\rm \AA}) = (\lambda_B/5100{\rm
\AA})^{(1-\alpha_{opt})} L_B$. More exact formula has recently been
provided by Greene \& Ho (2005). However, due to observational
uncertainties in estimating BLR parameters, both these expressions
give black hole masses comparable within errors.

\subsection{Seyfert galaxies and LINERs}
This sample contains objects selected from Ho \& Peng (2001) and  Ho
(2002; hereafter HP01 and H02, respectively). All objects studied by
HP01 have Seyfert 1 type nuclei and are taken from the Palomar and
CfA surveys. We selected  only those for which estimations of black
hole masses (obtained mostly by methods different form the virial
one discussed above) were available in the literature. The sample of
H02 is composed by AGNs with given  black hole masses and includes
Seyfert galaxies, LINERs, Transition Objects (Ho et al. 1993), and
PG quasars. We included only AGNs for which at least the H$\alpha$
line is broad, i.e., we did not include galaxies with nuclei of
spectral type S2, L2, and T2 (see Ho et al. 1997). We did not
include here PG quasars --- they are treated by us separately (see
\S2.4). The final 'Seyferts + LINERs' sample is made from 39 objects
(36 galaxies with Seyfert nuclei and 3 LINERs) listed in Table 3.
The table encloses information about IAU coordinates for the J2000.0
epoch; name of the source; distance of the source, $d$; absolute
$B$-magnitude of the nucleus, $M_B$; $B$-band luminosity of the
nucleus, $L_B$; total radio luminosity at $5$ GHz, $L_R$;
radio-loudness parameter, ${\cal R}$; black hole mass, ${\cal
M}_{\rm BH}$, followed by the appropriate reference; and finally the
$B$-band and radio luminosities in the Eddington units.

Distances, if less than 40 Mpc ($z < 0.009$), are the same  as
adopted by H02 from Tully (1988). They were derived taking into
account the effect of the Virgo infall. If larger, the values given
in H02 were multiplied by a factor $(0.75/0.7)$, due to the
difference in the value of the Hubble constant used in H02 and in
this paper. Absolute $B$-magnitudes of nuclei, $M_B$, were taken
from HP01 if available. These are the values calculated from
directly measured apparent magnitudes $m_B$ of the nuclear regions.
In other cases, $M_B$ were taken from H02. These are obtained from
the $L_{H\beta}-M_B$ correlation. For distances to the source larger
than 40 Mpc, the absolute magnitude $|M_B|$ was increased by a
factor $(2.5 \, \log [0.75/0.7]^2)$. The nuclear luminosity, $L_B$,
was then calculated from the absolute $B$-magnitude, $M_B$, using
the standard relation
\be \log L_B= 0.4 \, |M_B| + 35.6 \quad . \ee
Regarding the radio emission, contrary to HP01 and H02, we decided
to use in our studies of the AGN radio-loudness the total
luminosities, i.e. including nuclear and extended emission (see
\S3.4). The total radio luminosities are taken from H02 and HP01 and
as the optical luminosities they are corrected to account for
different values of the Hubble constant.

\subsection{FR I radio galaxies}
Our sample of FRI radio galaxies consists of objects which were
observed by Hubble Space Telescope and hence have determined optical
luminosities (or the appropriate upper limits) for their unresolved
cores (see Kharb \& Shastri 2004, and references therein) and in
addition estimated black hole masses (Cao \& Rawlings 2004; Woo \&
Urry 2002). Such sample contains 31 objects listed in Table 4 along
with the optical and radio data: the IAU coordinates for the J2000.0
epoch; name; redshift, $z$; $B$-band luminosity $L_B$, obtained
after recalculating $V$-band luminosities provided by Kharb \&
Shastri (2004) to the cosmology we are using and then converted to
$B$-band assuming optical spectral index $\alpha_{opt}=0.5$; total
radio flux at $5$ GHz, $F_5$, obtained from the literature indicated
in the table (and if originally provided at other frequencies,
recalculated assuming radio spectral index $\alpha_R=0.8$); total
$5$ GHz radio luminosity, $L_R$; radio-loudness parameter, ${\cal
R}$ (assuming that the accretion luminosity is equal to the one
observed from the optical core by HST); black hole mass, ${\cal
M}_{\rm BH}$, taken from Woo \& Urry (2002) and Cao \& Rawlings
(2004); and finally $B$-band and radio luminosities expressed in the
Eddington units.

\subsection{PG quasars}
This sample consists of those BQS objects (Schmidt \& Green 1983)
which have redshifts less than $z = 0.5$, black hole mass available
in literature, and are not included in our other sub-samples. The
BQS objects are commonly called PG (Palomar-Green) quasars, despite
the fact that not all of them satisfy the formal luminosity
criterion $M_V$ or $M_B < -23$ to be called quasars (there are 7 of
such BQS AGNs in our sample). In addition, many of the BQS sources
are classified as NLS1s (Narrow-Line Seyfert 1 Galaxies). These
objects represent high accretion-rate AGNs with relatively low black
hole masses. They are exceptionally radio-quiet as a class
(Ulvestad, Antonucci, \& Goodrich 1995; Greene, Ho, \& Ulvestad
2006), with only few reaching ${\cal R}> 100$, and none producing a
prominent extended radio structure  (Zhou \& Wang 2002; Whalen  et
al. 2006; Komossa et al. 2006). The final BQS sample is listed in
Table 5, along with the optical and radio data: the IAU coordinates
for the J2000.0 epoch; name; redshift, $z$; $B$-band luminosity
$L_B$, calculated for $m_B$ given by Schmidt \& Green (1983); $F_5$,
obtained from Kellerman et al. (1989); total $5$ GHz radio
luminosity, $L_R$; radio-loudness parameter, ${\cal R}$; black hole
mass, ${\cal M}_{\rm BH}$, taken from Vestergaard (2002) or Woo \&
Urry (2002); and finally $B$-band and radio luminosities expressed
in the Eddington units.

\section{Results}

\subsection{Global patterns}
Radio luminosities vs.  optical luminosities of the selected AGNs
are plotted  in Figure 1. As can be seen our sub-samples form two
sequences which are separated by $\sim 3$ orders of magnitude in
radio luminosity. One can check from Tables 1-5, that the upper
sequence is almost exclusively populated by objects with black hole
masses ${\cal M}_{\rm BH} > 10^8  {\cal M}_{\odot}$. The only
exception one object from the `Seyferts plus LINERs' sub-sample, NGC
1275. It should be noted, however, that this object, as all the
other AGNs in the upper sequence, is hosted by a giant elliptical
galaxy (specifically, by the cD galaxy of the Perseus cluster), and
has an extended FR I-like radio structure observed presumably at a
small angle to the line of sight (Pedlar et al. 1990). Close to the
upper sequence but still belonging to the 'Seyferts plus LINERs'
subsample is located NGC 4258 (M106). This spiral galaxy hosts an
extremely weak AGN. Its total radio luminosity is about 100 times
larger than the nuclear one, and most likely is not related to the
jet activity.

Whereas there are no disc-galaxy-hosted AGNs in the upper sequence,
AGNs hosted by giant elliptical galaxies are present in both the
upper and lower sequences. This particularly concerns highest
accretion luminosity objects, i.e. quasars. Most of them in fact,
even those with very massive black holes and resolved elliptical
hosts, occupy the lower sequence which we will call hereafter the
`radio-quiet sequence'(see \S{4.1}). At intermediate accretion
luminosities, AGNs hosted by giant elliptical galaxies and located
in the lower sequence are represented in our sample only by four
objects. However, recent discoveries of many radio-quiet galaxies
with very broad Balmer lines and hosting very massive black holes
(Strateva et al. 2003; Wu \& Liu 2004) strongly indicate that
rareness of such objects in the Eracleous \& Halpern samples might
be due only to selection effects. Hence, it is plausible that also
at intermediate accretion luminosities, most of AGNs hosted by giant
elliptical galaxies are radio-quiet.

An intriguing feature of the Figure 1 is that both the radio-quiet
sequence and the `radio-loud' upper sequence have a similar
dependence of radio-luminosity on the accretion-luminosity. It
corresponds to an approximate constancy of $L_R$ at larger values of
$L_B$ and to a decrease of $L_R$ at smaller accretion rates. As
shown in Figure 2, qualitatively the same feature is found when
luminosities are expressed in Eddington units. The main difference
between Figures 1 and 2 is the relative location of the two
sequences which reflect the fact that AGN black-holes in the
radio-loud sequence are on average $\sim 20$ times more massive than
black holes in the AGNs forming the radio-quiet sequence. In
Figure~2 this causes a left-down shift of the upper sequence
relative to the lower sequence. Obviously due to a wide range of
black hole masses in each of the sequences there are quantitative
differences in the location of individual objects within the
sub-samples. Furthermore, at the largest accretion luminosities,
where the lower pattern is occupied mostly by quasars hosted by
giant elliptical galaxies, the relative location of the two sequence
is not significantly modified.

In Figure 3, we plot the dependence of the radio-loudness, ${\cal
R}$, on the Eddington ratio, $\lambda$, assuming $\lambda =
L_{bol}/L_{Edd} = 10 \, (L_B /L_{Edd})$ (see, e.g., Richards et al.
2006).\footnote{Note that for very low luminosity AGNs the
bolometric correction can be a factor $\sim 2$ larger than
considered above. However, due to very large uncertainties and not
known functional dependence of the exact correction factor on the
luminosity (Ho 1999), we decided to use the same proportionality
constant for all the analyzed AGNs.} Our results confirm the trend
of the increase of radio-loudness with decreasing Eddington ratio,
originally noticed by Ho (2002; see also Merloni, Heinz, \& Di
Matteo 2003; Nagar, Falcke, \& Wilson 2005). However, we show in
addition that this trend is followed separately --- with a large
difference in normalization --- by the `radio-quiet' and the
`radio-loud' sequences. Yet another feature revealed in Figure 3 is
a clear change of slope of the ${\cal R}-\lambda$ dependence
indicating some sort of saturation of radio-loudness at low
Eddington ratios. A similar trend can be noticed, but specifically
for FR I and FR II radio galaxies, in the data presented by Zirbel
\& Baum (1995). Let us recall that almost all BLRGs and radio-loud
quasars in our samples have FR II radio morphology.

Finally, in Figure 4 we illustrate the dependence of radio-loudness
on black-hole mass.  This plot demonstrates that AGNs with the black
hole masses $> 10^8 {\cal M}_{\odot}$ reach values of radio-loudness
three orders of magnitude larger than AGNs with black hole masses
$<3 \times 10^7 {\cal M}_{\odot}$ on average\footnote{A number of
very radio-loud AGNs was claimed by Woo \& Urry (2002) to be
characterized by ${\cal M}_{\rm BH} < 10^8 \, {\cal M}_{\odot}$.
However, as it was demonstrated by Laor (2003), in most of these
cases the black hole masses have been determined incorrectly.}. A relatively
smooth transition between those two populations most likely is caused
by the overlap between black hole masses hosted by disc and elliptical
galaxies. Errors in black hole mass estimations can also have a similar
effect. It is interesting to compare our Figure 4 with the analogous
figures restricted to high Eddington-ratio objects presented by Laor
(2003) and McLure \& Jarvis (2004). One can see that in all cases
there is a difference of about $3$ orders of magnitude between the
maximal radio-loudness of AGNs with ${\cal M}_{\rm BH}/{\cal
M}_{\odot}> 10^8$ and AGNs with less massive black holes. However,
because in our sample we have included AGNs with very low
Eddington-ratios the boundaries of maximal radio-loudness for less
and more massive objects are now located at much larger ${\cal R}$.
This effect is a simple consequence of the radio-loudness increasing
with decreasing Eddington-ratio. Because of this, the upper radio
boundaries are determined by low-$\lambda$ objects: in the
lower-${\cal M}_{\rm BH}$ sub-group by Seyferts and LINERs, in the
larger-${\cal M}_{\rm BH}$ sub-group by FR Is.

\subsection{Incompleteness of our sample and related uncertainties}
Our sample is very heterogeneous, being composed from incomplete
sub-samples selected using different criteria. This must have
effects on the presented  plots and should be taken into account
when interpreting our results. The largest incompleteness concerns
the broad-line AGNs taken from the radio-selected samples. They
include BLRGs and radio-loud quasars, which are all associated with
giant elliptical galaxies. However, as known from radio studies of
optically selected quasars, the majority of such sources are
radio-quiet and many are radio-intermediate (White et al. 2000). The
same was recently found for lower-$\lambda$ objects, when
investigating the double-peaked broad emission lines in AGNs from
the SDSS catalog (Strateva et al. 2003). Contrary to the deeply
grounded conviction that the presence of double-peaked lines is
unique to BLRGs, it was discovered that such lines are pretty common
also in radio-quiet AGNs with black hole masses characteristic of
giant elliptical galaxies (Wu \& Liu 2004). Noting all that, one
should consider the upper, radio-loud sequence in our plots  as
populated only by a minority of the elliptical-hosted AGNs. In other
words, with a complete (though not yet available) sample, the
mid-Eddington AGNs in giant elliptical galaxies would not be
confined to the upper sequence, but would show continuous
distribution down to the lowest detectable radio levels, similarly
to the PG quasars in our sample.

Meanwhile, the results presented by Wu \& Liu (2004) in the right
panels of their Figure 1 seem to indicate that for $\lambda <
10^{-3}$ the proportions between radio-loud and radio-quiet
fractions discussed above can reverse, i.e., that for $\lambda <
10^{-3}$ the upper sequence is populated by the majority of the
elliptical-hosted AGNs. This seems to be confirmed by relatively
complete surveys of nearby galaxies, for which the prospect of
missing radio-quiet AGNs among giant elliptical galaxies is rather
low (see, e.g., Terashima \& Wilson 2005, Chiaberge et al. 2005).

\subsection{Potential errors and attempts to minimize them}

Some quantities used to construct  our plots can be subject to
significant  errors. This primarily concerns galaxies with very weak
nuclei. Sometimes it is even  difficult to decide whether the
observed nuclear features are dominated by the AGN or by starburst
activities, and thus what is the real accretion luminosity. To avoid
the impact of such uncertainties on our results we did not include
in the 'Seyferts + LINERs' sample objects of spectral type 2, i.e.
those with no broad lines. On the other hand, uncertainties concern
also sources with broad-line nuclei, firstly because of the
accretion-luminosity contamination by starlight, and secondly
because of the internal extinction. To avoid these uncertainties, Ho
(2002) calculated accretion luminosities indirectly, using the
correlation of the accretion luminosity (or more precisely --
absolute magnitude $M_B$) with the intensity of the H$\beta$ line.
However, as one can see in Ho \& Peng (2001), this correlation is
reasonable only for very luminous objects. Therefore, to minimize
the related uncertainties, for sources overlapped by samples from Ho
\& Peng (2001) and Ho (2002), we adopted nuclear luminosities
obtained in Ho \& Peng from the direct optical measurements and
corrected by subtracting the starlight.

Our selection of FR I radio-galaxies is less rigorous: in this
subsample  we included also the objects for which there are no
direct signatures of an accretion flow. We note that the correlation
between the optical fluxes of the nuclear cores observed in these
sources by the HST  and their radio fluxes suggests that the
observed nuclear emission is due to synchrotron radiation
originating in the inner portions of the jets (Verdoes Kleijn, Baum,
\& de Zeeuw 2002). Hence, in several papers it was assumed that HST
detections provide the upper limits for the optical radiation due to
the accretion flow (see, e.g., Chiaberge, Capetti, \& Celotti 1999;
Chiaberge, Capetti, \& Macchetto 2005). However, it does not have to
be the case if the central nuclei are hidden by ``dusty tori". Then
the situation might be the opposite: the accretion luminosity can be
in fact larger than the luminosity measured by HST. Arguments in
favor of such a possibility were put forward by Cao \& Rawlings
(2004), who postulated that indeed BL Lac objects accrete at the
high rates. There are, however, strict limits on the bolometric
luminosity of hidden AGN: it cannot be larger than the observed
infrared luminosity resulting from reprocessing of the hidden
nuclear radiation by the circumnuclear dust. Infrared observations
clearly indicate that active cores in FR I's radiate several orders
of magnitude below the Eddington level (Knapp, Bies, \& van Gorkom
1990; M\"uller et al. 2004; Haas et al. 2004). By comparing the
infrared luminosities with the HST measurements for a number of FR I
radio galaxies included in our sample, we find that the accretion
luminosities of some of them might be underestimated by a factor
$\gtrsim 10$.

Other errors which may significantly affect details of our plots
come  from the estimations of black-hole masses. For all the objects
except BLRGs and radio-loud quasars, these masses were taken from
the literature. They were obtained using variety of  methods.
However, by comparing Figure 1 and 2, which are constructed with and
without involving black hole masses, respectively, one can see that
such global features as two-sequence structure and the general
trends survive. Hence we conclude that our results are not
significantly affected by errors of black-hole mass estimations.

Finally, we should comment on our choice of calculating the
radio-loudness parameter by using not the strictly nuclear radio
fluxes (as in several other similar studies) but the total ones. In
the cases of lobe-dominated radio quasars, BLRGs, and FR I radio
galaxies, the observed radio luminosity is produced mainly by the
jet-powered extended radio structures. As long as our interest in
radio-loudness is to understand the strong diversity of the jet
power, the extended radio component should be therefore considered.
Less obvious is the situation in the case of AGNs hosted by spiral
galaxies. There the nuclear radio emission is usually dominant while
the extended radio emission can be due not only  to  jet activity,
but also to  starburst regions (but see Gallimore et al. 2006).
Hence by using total radio fluxes, as we did in this paper, one can
overestimate radio-loudness (which should concern only the
jet-related emission). Such a choice is however more appropriate
then taking into account only the nuclear radio component: in this
way we avoid underestimating the radio-loudness, which could have
the effect of increasing the gap between the upper and lower
patterns in Figures 1-3. In other words, our choice is conservative.
In this context we note that in eleven particular sources from our
`Seyferts + LINERs' subsamble (consisting of 39 objects), the total
radio luminosities are about twice as large as the core radio
luminosities identified to be related to the jet activity. Hence, in
these cases the radioloudness is presumably overestimated by a
factor of $\gtrsim 2$. This regards in particular NGC 1275, Ark 120,
Mrk 79, Mrk 110, NGC 3982, NGC 4203, NGC 4258, NGC 4565, NGC 4639,
Mrk 841, and NGC 5940.

\section{Discussion}

\subsection{Radio--loudness vs. galaxy morphology}
Using several subsamples of AGNs which together cover seven decades
in the Eddington ratio, we have demonstrated that radio-selected
AGNs hosted by giant elliptical galaxies can be about $10^3$ times
radio louder than AGNs hosted by disc galaxies, and that the
sequences formed by the two populations show the same trend --- an
increase of the radio-loudness with the decrease of the Eddington
ratio. This corresponds to a slower than linear decrease of the
(Eddington-scaled) radio luminosity with the decrease of the
Eddington-ratio. The same trend was discovered by Ho (2002), but
because his paper considered only nuclear radio luminosities and
didn't include  BLRG, the radio-loudness vs. Eddington-ratio plot
did not reveal a double-sequence structure.  A trend similar to ours
was found also by Terashima \& Wilson (2003), but in the plane
`radio-loudness vs. X-ray luminosity', where radio-loudness is
defined as the ratio of the radio to the X-ray luminosity. These
authors considered two cases, one with radio-loudness including only
nuclear radio luminosities and another with radio-loudness defined
for the total radio luminosity. In the latter case, in similarity to
our results, a double-sequence structure emerges. A double-sequence
structure was noticed also by Xu, Livio \& Baum (1999), in the
`radio-luminosity vs. [OIII] luminosity' plane. These results
strongly confirm that when one considers the total radio luminosity,
AGNs split into two sub-classes, with AGNs hosted by giant
ellipticals extending to much larger radio luminosities than AGNs
hosted by disk galaxies. It is the consequence of the fact that
radio-galaxies have powerful extended radio-emission that must be
obviously taken into account when estimating their radio fluxes.

The absence of prominent, extended radio structures in AGNs hosted
by disc galaxies, when confronted with giant elliptical hosts of the
classical double radio sources, led in the past to the conviction
that all AGNs in disc galaxies are radio-quiet, and that all AGNs in
giant elliptical galaxies are radio-loud. Only recently, after the
HST allowed to image and determine the host morphologies of distant,
luminous quasars, it has became clear that the claimed one-to-one
correspondence between the radio-loudness and the galaxy morphology
is invalid: a number of luminous, radio-quiet quasars have been
found to be hosted by giant elliptical galaxies (see Floyd et al.
2004 and references therein). This discovery stimulated in turn
searches for radio-loud AGNs among those hosted by disc galaxies.
Surely, Ho \& Peng (2001) demonstrated that after subtraction of the
starlight, nuclei of some Seyfert galaxies `become' radio-loud,
according to criterion ${\cal R}>10$ introduced by Kellerman (1989)
for quasars. The same was found for LINERs in nearby disc galaxies
(Ho 2002). However, as we have shown in this paper, all Seyfert
galaxies and LINERs remain well separated from the BLRGs and FR I
radio galaxies, and the fact that some of them reach ${\cal R}>10$
is caused by the increase of the radio-loudness with the decreasing
$\lambda$. Noting such a dependence, we propose to call an AGN
radio-loud if
\be \log {\cal R} > \log {\cal R}^* = \cases { -\log \lambda +1 &
for \,  $\log \lambda > -3$ \cr 4 & for \,  $\log \lambda < -3$ \cr
} \, . \label{rl} \ee

\subsection{Two sequences imply two parameters; comparison with BH
X-ray binaries}
Comparing observed properties of AGN to those of radio active
black-hole X-ray binaries provides important hints about the nature
(assumed to be common) of the jest production mechanism in these two
classes of objects. Gallo, Fender, \& Pooley (2003) discovered that
radio luminosities of BH binaries in low/hard states (i.e., at low
accretion rates) correlate with X-ray luminosities, $L_X$. They
found that luminosity variations of two objects, GX 339-4 and V404
Cyg, follow the relation $L_R \propto L_X^{0.7}$ which holds over
more than three order of magnitudes in $L_X$ with the same
normalization within a factor of 2.5. This discovery triggered
speculations that the powering of radio activity in XRBs during the
how/hard state is entirely determined by accretion. However, the
fact that a similar trend, albeit with a huge scatter, is followed
by both our AGN sequences (compare the low-$L_B/L_{\rm Edd}$
sections in our Figure 2 with Figure 2 from Gallo et al. [2003])
shows that radio-loudness cannot depend only on the accretion rate:
clearly an additional parameter is required to explain the bimodal
distribution of radio-loudness for spiral-hosted and
elliptical-hosted AGNs at low $\lambda$'s, and its significant
scatter within both sequences.

The monotonic dependence of radio luminosities on the Eddington
ratio in XRBs breaks down at $\lambda \sim 0.01$, and observations
clearly indicate the intermittency of the  jet production at higher
luminosities and its connection with a spectral state (Fender,
Belloni, \& Gallo 2004). Qualitatively a similar break but located
at smaller $\lambda$'s is seen in the AGN sequences.

Motivated by the similarities of the low- and high-$\lambda$
patterns in XRBs and AGNs, Merloni, Heinz, \& Di Matteo (2003)
proposed that at high accretion luminosities the jet production is
intermittent also in AGNs. This idea was recently explored and
supported by Nipoti, Blundell, \& Binney (2005). However, at high
accretion rates (just as at the low ones discussed previously)
spiral-hosted AGNs are located exclusively in the radio-quiet
sequence of our Figures 1-3, as opposed to quasars which populate
both upper and lower patterns. Only few exceptional cases are known
in which the high-$\lambda$ spiral-hosted AGNs reach radio-loudness
${\cal R}> 100$ (V\'eron-Cetti \& V\'eron 2001; Ledlow et al. 2001;
Whalen et al. 2006) --- all the other objects of this kind are $\sim
10^3$ times radio weaker than the radio-loud quasars and BLRGs with
similar values of $\lambda$. Clearly this fact indicates again that
an additional parameter must play a role in explaining why the
upper, radio-loud sequence in Figures 1-3 is reachable only by
the AGNs hosted by early type galaxies.

\subsection{The spin paradigm}
It is natural to expect that the second (in addition to
accretion rate) physical parameter determining radio-loudness of
AGNs is related to the properties of the central black-hole. In this
case we are left with only two possibilities: mass and spin.
Although the upper sequence is formed by massive black-holes only,
the mass cannot be the required physical parameter unless one
invents a specially designed mass-scaled jet launching mechanism.
The black-hole mass, however, reflects the history of its growth and
in this fashion  is related to both the galaxy's morphology and to
the other parameter: the spin. In the next section we will show how
the black-hole's spin can explain the radio-loudness  ``bimodality"
but since this idea has a rather long and eventful history, we will
first review shortly the past and present-day status of the
so-called `spin paradigm'.

In 1990, Blandford suggested that efficiency of the jet production
(assuming that a jet is powered by the rotating black hole via the
Blandford-Znajek (1977) mechanism) is determined by the black hole
spin, $J$, or, more precisely, by the dimensionless angular
momentum, $a \equiv J/J_{max}=c \, J/G \, {\cal M}_{\rm BH}^2$. This
is a very attractive because in principle it can explain the very
wide range of radio-loudness of AGNs that look very similar in many
other aspects. This spin paradigm was explored by Wilson \& Colbert
(1995) and Hughes \& Blandford (2003), who assumed that the spin
evolution is determined by black-hole mergers. They showed that
mergers of black holes which follow mergers of galaxies lead to a
broad, `bottom-heavy' distribution of the spin, consistent with a
distribution of radio-loudness in quasars. As shown by Moderski \&
Sikora (1996a,b) and Moderski, Sikora \& Lasota (1998),
'bottom-heavy' distribution of the spin is reachable also if the
evolution of the black hole spin is dominated by accretion, provided
the accretion history consists of many small accretion events with
randomly oriented angular momentum vectors. In such a case,
accretion events lead to formation of both co-rotating and
counter-rotating discs depending on the initial angular momenta,
which in turn spin-up and spin-down the black hole. This possibility
was questioned by Natarajan \& Pringle (1998) and Volonteri et al.
(2005) who  argued that angular momentum coupling between black
holes and accretion discs is so strong, that the innermost portions
of a disc are always forced to co-rotate with a black hole, and
therefore  all AGN black-holes should have large spins. However, as
was demonstrated by King et al. (2005) and confirmed by Lodato \&
Pringle (2006),  counter-rotating disks can in fact be formed,
provided individual accretion events involve much smaller amount of
material than the mass of the black hole.

\subsection{The revised spin paradigm}
Our new, revised version of the spin ``paradigm", based on the
existence of two radio-loudness sequences, states that the
radio-loud, upper-sequence AGNs have high black-hole spins. This
version of the spin (or rather ``spin-accretion") paradigm must be
completed by two elements:
\noindent
\renewcommand\theenumi{(\roman{enumi})}
\begin{enumerate}
\item Black hole evolution scenarios explaining why the spin
distribution of BHs hosted by giant elliptical galaxies extends to
much larger values than in the case of spiral galaxies;
\item A spin-accretion scenario explaining the intermittent jet
activity at high accretion rates.
\end{enumerate}

\noindent We will discuss the evolution problem first, but before
that let us just mention that in black-hole XRBs the situation is
simpler: since to reach the maximum spin a black hole has to double
its mass, black-hole spins in low-mass binaries do not evolve during
the lifetime of the systems. Therefore one should not expect in this
case the presence of two radio-loudness sequences as observed (Gallo
et al. 2003). Observations suggest `moderate' black-hole spin values
in several XRBs (see e.g. Shafee et al. 2006 and Davis, Done \&
Blaes 2006) so in principle they would be equivalent to the
``radio-loud" AGN sequence.

\subsubsection{Black-hole spin-up}

The spin-up of BHs in spiral galaxies can be limited by multiple
accretion events with random orientation of angular momentum
vectors, and small increments of accreted mass
\be m \ll m_{align} \sim a \sqrt{{R_S \over R_w}}{\cal M_{\rm BH}} \quad ,
\label{align} \ee
where $R_S=2GM/c^2$ is the Schwarzschild radius and $R_w (\sim 10^4
R_g)$ is the distance of the warp produced by the Bardeen-Petterson
process in the accretion disk, which at large distances is inclined
to the equatorial plane of the rotating black hole (Bardeen \&
Petterson 1975)\footnote{The required mass increments  may be even
smaller, if the viscosity related to the `vertical' shear is larger
than viscosity related to the `planar' shear and when the angular
momentum of the warped disk is smaller than the angular momentum of
the black hole (Papaloizou \& Pringle 1983; Natarajan \& Pringle
1998; see however discussion in King et al. 2005, Sect. 4.2).}. Only
with such small increments of accreting mass, a counter-rotating
accretion disk can survive over a whole accretion event, as
otherwise it would undergo a flip due to an alignment process which
tends to co-align a BH angular momentum with an angular momentum of
the distant ($R \ge R_w$) regions of the accretion disk (Rees 1978).
The accretion-event mass limit (Eq. \ref{align}) is severe but
consistent with observations indicating very short life-times of
individual accretion events in Seyfert galaxies (Capetti et al.
1999; Kharb et al. 2006) and showing a random orientation of jets
relative to the host galaxy axis (Kinney et al. 2000; Schmitt et al.
2001). It should be also noted that fueling of AGN in disk galaxies
is presumably not related to the galaxy mergers and is provided by
molecular clouds (Hopkins and Hernquist 2006). Alternatively, the
low values of spins of BHs in disk galaxies could be assured if the
BH growth is  dominated by mergers with intermediate mass BHs ---
relics of Population III stars or BHs formed in young stellar
clusters (see Mapelli, Ferrara, \& Rea 2006 and refs. therein).

In contrast to spiral galaxies, giant ellipticals underwent at least
one major merger in  the past (see, e.g., Hopkins et al. 2006). Such
mergers are followed by accretion events which involve too much mass
to satisfy the condition given by Eq.(\ref{align}). Then, regardless
of whether the disk was initially counter- or co-rotating, due to
the alignment process, all disks will co-rotate (counter-rotating
disks undergo flips). Provided that $m >> m_{align}$ they spin-up
black holes to large values of $a$, up to $a > 0.9$ if $m \sim
M_{BH}$. This is in agreement with the large average spin of quasars
as inferred from comparison between the local BH mass density and
the amount of radiation produced by luminous quasars (So{\l}tan
1982; Yu \& Tremaine 2002; Elvis, Risasliti, \& Zamorani 2002;
Marconi et al. 2004). This explains why AGNs on the upper branch
have high spins in full agreement with our version of the spin
paradigm. However, since most quasars are radio-quiet (a
``bottom-heavy" distribution of radio-loudness), the version of the
spin paradigm according to which the distribution of the
radio-loudness of quasars matches the distribution of their spin is
invalidated. This brings us to the second problem: the intermittent
jet activity at high accretion rates.

\subsubsection{Intermittency}

Despite the ``bottom heavy" distribution of quasar radio-loudness
one can still consider the spin to be the parameter which determines
the power of an outflow. However, such an outflow may fail to become
a jet unless it is collimated (Begelman \& Li 1994). Hence, we
suggest that the intermittent production of narrow jets, as seen
directly in XRBs, is related to intermittent collimation and that
the latter is provided by a surrounding non-relativistic MHD outflow
launched in the accretion disc. Such a double jet structure was
originally proposed by Sol, Pelletier, \& Asseo (1989), and shown to
provide good collimation by Bogovalov \& Tsinganos (2005), Gracia,
Tsinganos, \& Bogovalov (2005), and Beskin \& Nokhrina (2006).
Assuming that at high accretion rates the disc has two realizations,
being driven by viscous forces or by magnetic torques from MHD
winds, one can obtain intermittency of collimation by transitions
between two such accretion modes. As Livio, Pringle, \& King (2003)
suggested and Mayer \& Pringle (2006) confirmed, such transitions
can be governed by processes responsible for generation of a
poloidal magnetic field. An alternative scenario, proposed by Spruit
\& Uzdensky (2005), involves a drift of isolated patches of magnetic
fields to the center from very large distances.

At low $\lambda$'s intermittency is not observed (or rather it is
not deduced from observations) and as mentioned in Sect. 3.2 a ``top
heavy" distribution of radio-loudness for ellipticals is there
emerging.

\subsection{Other challenges?}

The spin paradigm may be challenged by the following observations:

\renewcommand\theenumi{(\roman{enumi})}
\begin{enumerate}
\item the very large spin of the central black-hole in the radio-quiet
Seyfert galaxy MCG-6-30-15 deduced from the profile of the
fluorescent iron line (Wilms et al. 2001);
\item the significant content of protons in quasar jets, as deduced from
the analysis of blazar spectra (Sikora \& Madejski 2000; Sikora et
al. 2005), and indicated by the discovery of circular polarization
in radio cores (Wardle et al. 1998; Homan, Attridge, \& Wardle
(2001);
\item the presence of relativistic jets in the neutron star
XRBs (Fomalont, Geldzahler \& Bradshaw 2001; Fender et al. 2004).
\end{enumerate}

\noindent
Can the  spin paradigm  be reconciled with these features?
\smallskip

\noindent {\it (i) The Seyfert galaxy MCG 6-30-15:} The
interpretation of the very extended and weak red wing of the
fluorescent iron line observed in the radio-quiet Seyfert galaxy MCG
6-30-15 in terms of the model involving rapid rotation of the
central black hole (Wilms et al. 2001) is not unique, and depends
crucially on the details of the continuum production in the vicinity
of the black hole. These details are very uncertain (Beckwith \&
Done 2004). Furthermore, broad fluorescent iron lines can be
produced also in the wind (Done \& Gierli\'nski 2006). Finally,
morphology of the host galaxy is E or SO (Ferruit, Wilson, \&
Mulchaey, 2000) and therefore the object could have undergone a
major merger. In addition, McHardy et al (2005) found $\lambda
\approx 0.4$ for this object. Hence, MCG 6-30-15 may indeed harbor a
rapidly spinning black hole, but as most \textsl{quasars}, it
happens to be observed in its radio quiet state.

\noindent
{\it (ii) Protons in quasar jets:} Theories of jet launching by rotating black holes predict a zero
proton content (see particular models by Hawley \& Krolik 2006;
McKinney 2006; and refs. therein). This is because proton loading
near the black hole is protected by the magnetic fields threading
the horizon (note that the Larmor radius of non-relativistic protons
is many orders of magnitude smaller than the size of an AGN jet
base). However, jets can be loaded by protons following an
interchange instabilities operating at the interface between the
central (relativistic) and the external (non-relativistic) outflows.
If both these outflows are Poynting flux dominated, then only the
non-relativistic one is efficiently self-collimated. Therefore the
two outflows press against each other resulting in mass exchange.
Additionally, both the components are subject to kink (and other
types of MHD) instabilities, which can increase the mass exchange
rate. Unfortunately, no quantitative model exists to show whether
efficient proton loading can be provided in this way. An alternative
possibility is that the central, relativistic outflow is launched by
the innermost portions of an accretion disc. Since the disc's
parameters in the very central region depend strongly on the black
hole spin, it is possible that the dependence of a jet power on the
black hole spin is simply due to larger efficiency of a jet
production in the case of a matter accreting onto rapidly rotating
black hole. As pointed out by Sikora et al. (2005), also in this
case the central portion of a jet can be relativistic, provided
magnetic field lines are at small angles to the disc axis.

\noindent
{\it (iii) Neutron star XRBs:} The discovery of relativistic jets in XRB systems with neutron stars
proves that the existence of an ergosphere, which plays a key role
in an extraction of a black hole rotational energy, is not necessary
for producing relativistic jets. Does it contradict the proposed
spin paradigm? Let us recall that the condition for launching a
Poynting-flux dominated outflow, which later becomes converted to a
matter dominated relativistic jet, is to supply a high
magnetic--to--rest-mass energy ratio ($\gg 1$) at the base of the
outflow. But this condition is obviously satisfied in the case of
the magnetic field anchored on a neutron star. Moreover, as in the
black-hole systems, slower winds from the accretion disc can
collimate the central portion of the outflow launched from the
rotating neutron star. Such a scenario provides a natural
explanation for an abrupt drop of jet radio luminosities in neutron
star XRBs below a certain X-ray luminosity (see Fig.3 in Migliari \&
Fender 2006): since the magnetic field anchored on a neutron star
does not depend on the accretion rate, while the magnetic field in
the disc does, the collimation of the outflow from a neutron star by
a wind from the disc breaks below a given accretion rate. Of course,
the lower the neutron-star magnetic field, the lower the accretion
luminosity at the break, which can explain why millisecond accreting
XRB pulsars are detected in radio even at very low accretion
luminosities (see Fig.2 in Migliari \& Fender 2006).

\section{Conclusions}
The main conclusions of our studies are:
\begin{itemize}
\item the upper boundaries of radio-loudness of AGNs hosted by giant elliptical
galaxies are by $\sim 3$ orders of magnitude larger than upper boundaries
of radio-loudness of AGNs hosted by disc galaxies;
\item both populations of spiral-hosted and elliptical-hosted AGNs
show a similar dependence of the upper bounds of the radio loudness
parameter ${\cal R}$ on the Eddington ratio $\lambda$: the radio
loudness increases with decreasing Eddington ratio, faster at higher
accretion rates, slower at lower accretion rates;
\item the huge, host-morphology-related difference between the radio-loudness
reachable by AGNs in disc and elliptical galaxies can be explained by
the scenario according to which
\renewcommand\theenumi{(\roman{enumi})}
\begin{enumerate}
\item the spin of a black hole determines the outflow's
power (in neutron star XRBs it is  the neutron star rotation),
\item central black holes can reach large spins only in early type galaxies
(following major mergers), and not (in a statistical sense) in
spiral galaxies;
\end{enumerate}
\item a broad, ``bottom-heavy'' distribution of radio-loudness in quasars
is not related to the distribution of the spin; however, it is still
the BH spin which mediates launching of the jet and determines upper
bound on the radio-loudness, whereas the interruptions in the jet
production is suggested to be caused by intermittency of a jet
collimation by MHD winds from  the accretion disk.
\end{itemize}

\section*{Acknowledgments}
M.S. was partially supported by Polish MEiN grant 1 P03D 00928 and
in 2005 by a CNRS ``poste rouge" at IAP. {\L}.S. acknowledges
support by the MEiN grant 1 P03D 00329 and by the ENIGMA Network
through the grant HPRN-CT-2002-00321. {\L}.S. thanks the Fellows of
the IAP for their hospitality and support during his stays there.
JPL was supported in part by a grant from the CNES. He thanks the
participants of the KITP program ``Physics of Galactic Nuclei" for
helpful remarks on the first version of this paper. We thank Greg
Madejski, Jack Sulentic, Ari Laor, and the anonymous ApJ referee for
helpful comments and suggestions. This research was supported in
part by the National Science Foundation under Grant No. PHY99-07949
and by the Department of Energy contract to SLAC no.
DE-AC3-76SF00515.

\clearpage

\begin{deluxetable}{lllllllllllllll}
\tabletypesize{\scriptsize}
\rotate
\tablecaption{Broad-Line Radio Galaxies}
\tablewidth{0pt}
\tablehead{
\colhead{IAU} & \colhead{name} & \colhead{$z$} & \colhead{$m_V$} & \colhead{$A_V$} & \colhead{$\kappa_{\star}$} & \colhead{$\log L_B$} & \colhead{$F_5$} & \colhead{ref.} & \colhead{$\log L_R$} & \colhead{$\log {\cal R}$} & \colhead{$FWHM_{H\alpha}$} & \colhead{$\log {\cal M}_{\rm BH}$} & \colhead{$\log L_B$} & \colhead{$\log L_R$}\\ J2000.0 & & & & & & [erg/s] & [Jy] & to col. 8 & [erg/s] & & [km/s] & [${\cal M}_{\odot}$] & [$L_{Edd}$] & [$L_{Edd}$]\\
(1) & (2) & (3) & (4) & (5) & (6) & (7) & (8) & (9) & (10) & (11) & (12) & (13) & (14) & (15)}
\startdata
0038-0207 & 3C 17 & 0.22 & 18 & 0.08 & 0.58 & 43.9 & 2.48 & 1 & 43.2 & 4.45 & 11500 & 8.7 & -2.9 & -3.6\\
0044+1211 & 4C +11.06 & 0.226 & 19 & 0.26 & 0.28 & 43.8 & 0.22 & 1 & 42.2 & 3.49 & 4299 & 7.8 & -2.1 & -3.7\\
0207+2931 & 3C 59 & 0.11 & 16 & 0.21 & 0.28 & 44.4 & 0.67 & 2 & 42.0 & 2.78 & 9800 & 8.9 & -2.6 & -5.0\\
0224+2750 & 3C 67 & 0.311 & 18.6 & 0.42 & 0.82 & 43.8 & 0.87 & 2 & 43.1 & 4.48 & 6200 & 8.1 & -2.4 & -3.1\\
0238-3048 & IRAS 02366-3101 & 0.062 & 15 & 0.22 & 0.3 & 44.2 & 0.00343 & 3 & 39.2 & 0.10 & 7800 & 8.6 & -2.5 & -7.5\\
0238+0233 & PKS 0236+02 & 0.207 & 17.7 & 0.11 & 0.46 & 44.1 & 0.12 & 1 & 41.9 & 2.89 & 11200 & 8.8 & -2.8 & -5.1\\
0312+3916 & B2 0309+39 & 0.161 & 18.2 & 0.49 & 0.1 & 44.0 & 0.822 & 2 & 42.5 & 3.55 & 6300 & 8.3 & -2.4 & -3.9\\
0342-3703 & PKS 0340-37 & 0.285 & 18.6 & 0.03 & 0.19 & 44.2 & 0.71 & 1 & 42.9 & 3.89 & 9800 & 8.8 & -2.7 & -3.9\\
0343+0457 & 3C 93 & 0.357 & 19.2 & 0.8 & 0.43 & 44.3 & 0.91 & 1 & 43.3 & 4.09 & 19600 & 9.5 & -3.3 & -4.3\\
0452-1812 & MS 0450.3-1817 & 0.059 & 17.8 & 0.14 & 0.9 & 42.2 & 0.0026 & 4 & 39.0 & 1.97 & 10900 & 7.5 & -3.4 & -6.6\\
0519-4546 & Pictor A & 0.035 & 16.2 & 0.14 & 0.14 & 43.3 & 15.54 & 5 & 42.3 & 4.17 & 18400 & 8.7 & -3.5 & -4.5\\
0832+3707 & CBS 74 & 0.092 & 16 & 0.12 & 0.17 & 44.2 & 0.00424 & 6 & 39.6 & 0.56 & 9200 & 8.7 & -2.6 & -7.2\\
0849+0949 & PKS 0846+101 & 0.365 & 19.2 & 0.19 & 0.11 & 44.3 & 0.1 & 1 & 42.3 & 3.18 & 9600 & 8.8 & -2.7 & -4.6\\
0859-1922 & PKS 0857-191 & 0.361 & 19.7 & 0.69 & 0 & 44.3 & 0.4 & 1 & 42.9 & 3.73 & 6500 & 8.5 & -2.3 & -3.7\\
0914+0507 & 4C +05.38 & 0.302 & 17.4 & 0.17 & 0.51 & 44.6 & 0.22 & 1 & 42.5 & 3.06 & 10600 & 9.1 & -2.7 & -4.7\\
0923-2135 & PKS 0921-213 & 0.053 & 16.5 & 0.2 & 0.65 & 43.2 & 0.42 & 1 & 41.1 & 3.09 & 8300 & 7.9 & -2.9 & -4.9\\
0947+0725 & 3C 227 & 0.086 & 16.3 & 0.09 & 0.4 & 43.9 & 2.6 & 1 & 42.4 & 3.62 & 13900 & 8.9 & -3.1 & -4.6\\
1030+3102 & B2 1028+31 & 0.178 & 16.7 & 0.27 & 0.18 & 44.6 & 0.172 & 2 & 41.9 & 2.40 & 6400 & 8.7 & -2.2 & -4.9\\
1154-3505 & PKS 1151-34 & 0.258 & 17.8 & 0.28 & 0.75 & 44.0 & 2.74 & 1 & 43.4 & 4.56 & 13400 & 8.9 & -3.0 & -3.6\\
1257-3334 & PKS 1254-333 & 0.19 & 18.6 & 0.28 & 0.22 & 43.9 & 0.54 & 1 & 42.4 & 3.68 & 6300 & 8.2 & -2.4 & -3.9\\
1332+0200 & 3C 287.1 & 0.216 & 18.3 & 0.08 & 0.36 & 44.0 & 1.35 & 1 & 43.0 & 4.12 & 4700 & 8.0 & -2.1 & -3.1\\
1407+2827 & Mrk 0668 & 0.077 & 15.4 & 0.06 & 0.27 & 44.2 & 2.421 & 2 & 42.2 & 3.15 & 6000 & 8.4 & -2.3 & -4.2\\
1419-1928 & PKS 1417-19 & 0.12 & 16.7 & 0.28 & 0.05 & 44.3 & 0.83 & 1 & 42.2 & 3.01 & 4900 & 8.3 & -2.1 & -4.2\\
1443+5201 & 3C 303 & 0.141 & 17.3 & 0.06 & 0.73 & 43.6 & 1.044 & 2 & 42.4 & 3.99 & 6800 & 8.0 & -2.6 & -3.7\\
1516+0015 & PKS 1514+00 & 0.053 & 15.6 & 0.18 & 0.76 & 43.4 & 1.37 & 1 & 41.7 & 3.42 & 4300 & 7.5 & -2.2 & -3.9\\
1533+3544 & 4C +35.37 & 0.157 & 17.8 & 0.08 & 0 & 44.1 & 0.129 & 2 & 41.6 & 2.70 & 4300 & 8.0 & -2.0 & -4.5\\
1617-3222 & 3C 332 & 0.151 & 16 & 0.08 & 0.85 & 43.9 & 0.92 & 2 & 42.4 & 3.66 & 23200 & 9.3 & -3.5 & -5.0\\
1637+1149 & MC2 1635+119 & 0.147 & 16.5 & 0.17 & 0.14 & 44.5 & 0.051 & 2 & 41.2 & 1.81 & 4900 & 8.4 & -2.0 & -5.3\\
1719+4858 & Arp 102B & 0.024 & 14.8 & 0.08 & 0.86 & 42.7 & 0.159 & 2 & 40.0 & 2.43 & 16000 & 8.2 & -3.6 & -6.3\\
1742+1827 & PKS 1739+184 & 0.186 & 17.5 & 0.21 & 0.11 & 44.3 & 0.39 & 1 & 42.3 & 3.06 & 13600 & 9.2 & -2.9 & -5.0\\
1835+3241 & 3C 382 & 0.058 & 15.4 & 0.23 & 0.06 & 44.1 & 2.281 & 2 & 42.0 & 2.95 & 11800 & 8.9 & -2.9 & -5.1\\
1842+7946 & 3C 390.3 & 0.056 & 15.4 & 0.24 & 0.31 & 44.0 & 4.45 & 5 & 42.2 & 3.37 & 11900 & 8.8 & -2.9 & -4.7\\
2101-4219 & PKS 2058-425 & 0.223 & 17.2 & 0.13 & 0 & 44.6 & 0.71 & 1 & 42.7 & 3.19 & 4600 & 8.4 & -1.9 & -3.8\\
2223-0206 & 3C 445 & 0.056 & 15.8 & 0.27 & 0.33 & 43.8 & 2.12 & 1 & 41.9 & 3.20 & 5600 & 8.0 & -2.3 & -4.3\\
2303-1841 & PKS 2300-18 & 0.129 & 17.8 & 0.11 & 0.14 & 43.8 & 0.89 & 1 & 42.3 & 3.59 & 8700 & 8.4 & -2.7 & -4.2\\
2307+1901 & PKS 2305+188 & 0.313 & 17.5 & 0.44 & 0.56 & 44.6 & 0.44 & 1 & 42.8 & 3.34 & 4400 & 8.4 & -1.9 & -3.7\\
2330+1702 & MC3 2328+167 & 0.28 & 18.3 & 0.15 & 0.37 & 44.2 & 0.078 & 2 & 42.0 & 2.87 & 3200 & 7.8 & -1.7 & -4.0\\
\enddata
\tablecomments{
Reference in column 9: [1] Wright \& Otrupcek (1990), [2] Gregory \& Condon (1991), [3] Condon et al. (1998), [4] Feigelson et al. (1982), [5] Kuhr et al. (1981), [6] Becker et al. (1995), [7] Wright et al. (1994), [8] Becker et al. (1991).}
\end{deluxetable}

\clearpage

\begin{deluxetable}{lllllllllllllll}
\tabletypesize{\scriptsize}
\rotate
\tablecaption{Radio-Loud Quasars}
\tablewidth{0pt}
\tablehead{
\colhead{IAU} & \colhead{name} & \colhead{$z$} & \colhead{$m_V$} & \colhead{$A_V$} & \colhead{$\kappa_{\star}$} & \colhead{$\log L_B$} & \colhead{$F_5$} & \colhead{ref.} & \colhead{$\log L_R$} & \colhead{$\log {\cal R}$} & \colhead{$FWHM_{H\alpha}$} & \colhead{$\log {\cal M}_{\rm BH}$} & \colhead{$\log L_B$} & \colhead{$\log L_R$}\\ J2000.0 & & & & & & [erg/s] & [Jy] & to col. 8 & [erg/s] & & [km/s] & [${\cal M}_{\odot}$] & [$L_{Edd}$] & [$L_{Edd}$]\\
(1) & (2) & (3) & (4) & (5) & (6) & (7) & (8) & (9) & (10) & (11) & (12) & (13) & (14) & (15)}
\startdata
0019+2602 & 4C 25.01 & 0.284 & 15.4 & 0.1 & 0 & 45.6 & 0.405 & 2 & 42.7 & 2.24 & 4600 & 9.1 & -1.6 & -4.5\\
0113+2958 & B2 0110+29 & 0.363 & 17 & 0.21 & 0 & 45.2 & 0.311 & 2 & 42.8 & 2.73 & 7200 & 9.2 & -2.1 & -4.5\\
0157+3154 & 4C 31.06 & 0.373 & 18 & 0.18 & 0.11 & 44.8 & 0.394 & 2 & 43.0 & 3.30 & 9000 & 9.1 & -2.4 & -4.3\\
0202-7620 & PKS 0202-76 & 0.389 & 16.9 & 0.17 & 0 & 45.3 & 0.8 & 1 & 43.3 & 3.12 & 6400 & 9.2 & -2.0 & -4.0\\
0217+1104 & PKS 0214+10 & 0.408 & 17 & 0.36 & 0.01 & 45.4 & 0.46 & 1 & 43.1 & 2.85 & 4500 & 8.9 & -1.7 & -3.9\\
0311-7651 & PKS 0312-77 & 0.225 & 16.1 & 0.32 & 0 & 45.2 & 0.59 & 1 & 42.6 & 2.59 & 2900 & 8.4 & -1.3 & -3.9\\
0418+3801 & 3C 111 & 0.049 & 18 & 5.46 & 0.04 & 45.1 & 6.637 & 2 & 42.3 & 2.35 & 4800 & 8.8 & -1.8 & -4.6\\
0559-5026 & PKS 0558-504 & 0.138 & 15 & 0.15 & 0 & 45.1 & 0.121 & 7 & 41.5 & 1.52 & 1000 & 7.4 & -0.4 & -4.1\\
0745+3142 & B2 0742+31 & 0.462 & 16 & 0.23 & 0 & 45.9 & 0.957 & 2 & 43.5 & 2.82 & 6500 & 9.6 & -1.8 & -4.2\\
0815+0155 & PKS 0812+02 & 0.402 & 17.1 & 0.1 & 0.01 & 45.2 & 0.77 & 1 & 43.3 & 3.22 & 4600 & 8.8 & -1.7 & -3.6\\
0839-1214 & 3C 206 & 0.197 & 15.8 & 0.15 & 0 & 45.1 & 0.72 & 1 & 42.6 & 2.63 & 5100 & 8.8 & -1.9 & -4.4\\
0927-2034 & PKS 0925-203 & 0.347 & 16.4 & 0.19 & 0 & 45.4 & 0.7 & 1 & 43.1 & 2.85 & 2200 & 8.3 & -1.0 & -3.3\\
0954+0929 & 4C +09.35 & 0.298 & 17.2 & 0.11 & 0 & 44.9 & 0.18 & 1 & 42.4 & 2.61 & 5100 & 8.7 & -1.9 & -4.4\\
1006-4136 & PKS 1004-217 & 0.33 & 16.9 & 0.2 & 0 & 45.2 & 0.3 & 1 & 42.7 & 2.68 & 2100 & 8.1 & -1.1 & -3.5\\
1007+1248 & PKS 1004+13 & 0.24 & 15.2 & 0.13 & 0 & 45.5 & 0.42 & 1 & 42.5 & 2.16 & 6100 & 9.3 & -1.9 & -4.9\\
1013-2831 & PKS 1011-282 & 0.255 & 16.9 & 0.21 & 0 & 44.9 & 0.29 & 1 & 42.4 & 2.65 & 4100 & 8.5 & -1.7 & -4.2\\
1022-1037 & PKS 1020-103 & 0.197 & 16.1 & 0.15 & 0 & 45.0 & 0.49 & 1 & 42.4 & 2.58 & 8700 & 9.2 & -2.4 & -4.9\\
1051-0918 & 3C 246 & 0.345 & 16.8 & 0.14 & 0 & 45.2 & 0.7 & 1 & 43.1 & 3.03 & 6300 & 9.1 & -2.0 & -4.1\\
1103-3251 & PKS 1101-325 & 0.355 & 16.5 & 0.31 & 0 & 45.4 & 0.73 & 2 & 43.2 & 2.86 & 3500 & 8.8 & -1.4 & -3.7\\
1107+3616 & B2 1104+36 & 0.392 & 18 & 0.06 & 0 & 44.8 & 0.217 & 2 & 42.7 & 3.04 & 6800 & 8.9 & -2.2 & -4.3\\
1131+3114 & B2 1128+31 & 0.29 & 16 & 0.07 & 0.01 & 45.3 & 0.31 & 1 & 42.6 & 2.39 & 4000 & 8.8 & -1.6 & -4.3\\
1148-0404 & PKS 1146-037 & 0.341 & 16.9 & 0.11 & 0.5 & 44.9 & 0.34 & 2 & 42.8 & 3.07 & 5000 & 8.7 & -1.9 & -4.0\\
1153+4931 & LB 2136 & 0.333 & 17.1 & 0.07 & 0 & 45.0 & 0.702 & 2 & 43.1 & 3.18 & 4400 & 8.7 & -1.7 & -3.7\\
1159+2106 & TXS 1156+213 & 0.347 & 17.5 & 0.09 & 0.43 & 44.7 & 0.085 & 2 & 42.2 & 2.66 & 7600 & 8.9 & -2.3 & -4.8\\
1210+3157 & B2 1208+32A & 0.389 & 16.7 & 0.06 & 0 & 45.3 & 0.16 & 2 & 42.6 & 2.39 & 5900 & 9.1 & -1.9 & -4.7\\
1225+2458 & B2 1223+25 & 0.268 & 17.1 & 0.07 & 0 & 44.8 & 0.138 & 2 & 42.2 & 2.47 & 5400 & 8.7 & -2.0 & -4.7\\
1235-2512 & PKS 1233-24 & 0.355 & 17.2 & 0.32 & 0 & 45.2 & 0.61 & 1 & 43.1 & 3.06 & 4900 & 8.9 & -1.8 & -3.9\\
1252+5624 & 3C 277.1 & 0.32 & 17.9 & 0.04 & 0 & 44.7 & 0.883 & 2 & 43.1 & 3.61 & 3200 & 8.1 & -1.6 & -3.1\\
1305-1033 & PKS 1302-102 & 0.278 & 15.2 & 0.14 & 0 & 45.7 & 1 & 1 & 43.1 & 2.54 & 3400 & 8.9 & -1.3 & -3.9\\
1349-1132 & PKS 1346-112 & 0.341 & 18 & 0.21 & 0 & 44.8 & 0.58 & 1 & 43.0 & 3.40 & 2300 & 7.9 & -1.3 & -3.0\\
1353+2631 & B2 1351+26 & 0.308 & 17.2 & 0.05 & 0.1 & 44.9 & 0.098 & 2 & 42.2 & 2.42 & 7800 & 9.1 & -2.3 & -5.0\\
1359-4152 & PKS 1355-41 & 0.314 & 15.9 & 0.29 & 0.05 & 45.5 & 1.4 & 1 & 43.3 & 2.93 & 9800 & 9.7 & -2.3 & -4.5\\
1423-5055 & CSO 0643 & 0.276 & 16.7 & 0.04 & 0.28 & 44.9 & 0.225 & 2 & 42.4 & 2.68 & 9000 & 9.2 & -2.4 & -4.9\\
1454-3747 & PKS 1451-375 & 0.314 & 16.7 & 0.26 & 0 & 45.2 & 1.84 & 1 & 43.4 & 3.36 & 3800 & 8.7 & -1.6 & -3.3\\
1514+3650 & 4C +37.43 & 0.371 & 16.3 & 0.07 & 0.04 & 45.4 & 0.361 & 2 & 42.9 & 2.59 & 7500 & 9.4 & -2.1 & -4.6\\
1527+2233 & LB 9743 & 0.254 & 16.7 & 0.18 & 0 & 45.0 & 0.16 & 2 & 42.2 & 2.33 & 3700 & 8.5 & -1.6 & -4.4\\
1609+1756 & 4C + 18.47 & 0.346 & 18 & 0.18 & 0 & 44.8 & 0.28 & 1 & 42.7 & 3.10 & 6100 & 8.8 & -2.1 & -4.2\\
1704+6044 & 3C 351 & 0.372 & 15.3 & 0.08 & 0 & 45.9 & 1.258 & 2 & 43.5 & 2.71 & 13000 & 10.2 & -2.4 & -4.9\\
1721+3542 & B2 1719+35 & 0.283 & 17.5 & 0.14 & 0.14 & 44.7 & 0.877 & 2 & 43.0 & 3.47 & 6000 & 8.7 & -2.1 & -3.8\\
1723+3417 & B2 1721+34 & 0.205 & 16.5 & 0.12 & 0 & 44.8 & 0.65 & 2 & 42.6 & 2.87 & 2300 & 8.0 & -1.2 & -3.5\\
1728+0427 & PKS 1725+044 & 0.297 & 17 & 0.47 & 0 & 45.1 & 1.21 & 1 & 43.2 & 3.21 & 3300 & 8.5 & -1.5 & -3.4\\
1748+1619 & MRC 1745+163 & 0.392 & 17.6 & 0.31 & 0 & 45.1 & 0.146 & 2 & 42.6 & 2.61 & 4200 & 8.7 & -1.7 & -4.2\\
1917-4530 & PKS 1914-45 & 0.364 & 16.8 & 0.27 & 0.1 & 45.3 & 0.18 & 1 & 42.6 & 2.44 & 9800 & 9.5 & -2.4 & -5.1\\
2142-0437 & PKS 2140-048 & 0.345 & 18 & 0.11 & 0 & 44.7 & 0.6 & 1 & 43.1 & 3.46 & 3900 & 8.4 & -1.7 & -3.4\\
2143+1743 & OX +169 & 0.211 & 15.7 & 0.37 & 0.01 & 45.3 & 1.061 & 8 & 42.8 & 2.67 & 4000 & 8.8 & -1.6 & -4.1\\
2211-1328 & PKS 2208-137 & 0.391 & 17 & 0.15 & 0 & 45.3 & 0.53 & 1 & 43.1 & 2.99 & 4100 & 8.8 & -1.6 & -3.8\\
2230-3942 & PKS 2227-399 & 0.318 & 17.9 & 0.06 & 0 & 44.7 & 1.02 & 1 & 43.2 & 3.67 & 6700 & 8.8 & -2.2 & -3.7\\
2250+1419 & PKS 2247+14 & 0.235 & 16.9 & 0.17 & 0 & 44.8 & 1.11 & 1 & 42.9 & 3.25 & 3500 & 8.3 & -1.6 & -3.5\\
2305-7103 & PKS 2302-713 & 0.384 & 17.5 & 0.1 & 0 & 45.0 & 0.15 & 1 & 42.6 & 2.66 & 4600 & 8.7 & -1.8 & -4.3\\
2351-0109 & PKS 2349-01 & 0.174 & 15.3 & 0.09 & 0.05 & 45.1 & 0.68 & 1 & 42.4 & 2.44 & 5800 & 9.0 & -2.0 & -4.6\\
\enddata
\tablecomments{
Reference in column 9: [1] Wright \& Otrupcek (1990), [2] Gregory \& Condon (1991), [3] Condon et al. (1998), [4] Feigelson et al. (1982), [5] Kuhr et al. (1981), [6] Becker et al. (1995), [7] Wright et al. (1994), [8] Becker et al. (1991).}
\end{deluxetable}

\clearpage

\begin{deluxetable}{lllllllllll}
\tabletypesize{\scriptsize}
%\rotate
\tablecaption{Seyfert Galaxies and Liners}
\tablewidth{0pt}
\tablehead{
\colhead{IAU} & \colhead{name} & \colhead{$d$} & \colhead{$|M_B|$} & \colhead{$\log L_B$} & \colhead{$\log L_R$} & \colhead{$\log {\cal R}$} & \colhead{$\log {\cal M}_{\rm BH}$} & \colhead{ref.} & \colhead{$\log L_B$} & \colhead{$\log L_R$}\\ J2000.0 & & [Mpc] & & [erg/s] & [erg/s] & & [${\cal M}_{\odot}$] & to col. 8 & [$L_{Edd}$] & [$L_{Edd}$]\\
(1) & (2) & (3) & (4) & (5) & (6) & (7) & (8) & (9) & (10) & (11)}
\startdata
0006+2012 & Mrk 335 & 114.2 & 18.33 & 42.9 & 38.4 & 0.58 & 6.8 & 1 & -2.0 & -6.6\\
0123-5848 & Fairall 9 & 214.1 & 23.28 & 44.9 & 39.7 & -0.04 & 7.9 & 1 & -1.1 & -6.3\\
0214-0046 & Mrk 590 & 117.0 & 16.61 & 42.2 & 39.2 & 2.07 & 7.3 & 1 & -3.1 & -6.2\\
0242-0000 & NGC 1068 & 14.4 & 16.47 & 42.2 & 39.5 & 2.46 & 7.2 & 1 & -3.2 & -5.8\\
0319+4130 & NGC 1275 & 75.1 & 18.68 & 43.1 & 42.2 & 4.30 & 8.5 & 2 & -3.6 & -4.4\\
0516-0008 & Ark 120 & 144.2 & 22.77 & 44.7 & 39.2 & -0.36 & 8.3 & 1 & -1.7 & -7.2\\
0742+4948 & Mrk 79 & 97.8 & 20.08 & 43.6 & 38.8 & 0.35 & 7.7 & 1 & -2.2 & -7.0\\
0919+6912 & NGC 2787 & 7.5 & 8.27 & 38.9 & 36.4 & 2.62 & 7.6 & 1 & -6.8 & -9.3\\
0925+5217 & Mrk 110 & 158.3 & 19.55 & 43.4 & 38.9 & 0.64 & 6.7 & 1 & -1.5 & -6.0\\
0955+6903 & NGC 3031 & 3.9 & 11.73 & 40.3 & 37.1 & 2.01 & 7.8 & 1 & -5.6 & -8.8\\
1023+1951 & NGC 3227 & 20.6 & 16.01 & 42.0 & 38.0 & 1.16 & 7.6 & 1 & -3.7 & -7.7\\
1106+7234 & NGC 3516 & 38.9 & 17.21 & 42.5 & 38.0 & 0.71 & 7.4 & 1 & -3.0 & -7.4\\
1139+3154 & Mrk 744 & 41.6 & 17.56 & 42.6 & 37.9 & 0.45 & 7.5 & 3 & -3.0 & -7.7\\
1139-3744 & NGC 3783 & 38.5 & 19.01 & 43.2 & 38.4 & 0.34 & 7.0 & -- & -1.9 & -6.7\\
1156+5507 & NGC 3982 & 17.0 & 11.76 & 40.3 & 37.7 & 2.53 & 6.1 & 2 & -3.9 & -6.5\\
1157+5527 & NGC 3998 & 14.1 & 12.95 & 40.8 & 38.0 & 2.42 & 8.7 & 1 & -6.1 & -8.8\\
1203+4431 & NGC 4051 & 17.0 & 14.97 & 41.6 & 37.4 & 0.93 & 6.1 & 1 & -2.7 & -6.9\\
1210+3924 & NGC 4151 & 20.3 & 19.18 & 43.3 & 38.5 & 0.43 & 7.2 & 1 & -2.0 & -6.7\\
1215+3311 & NGC 4203 & 14.1 & 10.58 & 39.8 & 36.7 & 2.00 & 7.9 & 1 & -6.2 & -9.3\\
1218+2948 & Mrk 766 & 55.4 & 16.72 & 42.3 & 38.4 & 1.32 & 6.6 & 4 & -2.4 & -6.3\\
1218+4718 & NGC 4258 & 7.3 & 8.17 & 38.8 & 38.0 & 4.34 & 7.6 & 1 & -6.9 & -7.7\\
1225+1239 & NGC 4388 & 16.8 & 13.17 & 40.8 & 38.0 & 2.33 & 6.8 & 4 & -4.1 & -6.9\\
1236+2559 & NGC 4565 & 9.7 & 10.19 & 39.7 & 37.6 & 3.04 & 7.7 & 4 & -6.2 & -8.3\\
1237+1149 & NGC 4579 & 16.8 & 12.81 & 40.7 & 38.0 & 2.45 & 7.9 & 4 & -5.3 & -8.0\\
1239-0520 & NGC 4593 & 39.5 & 17.8 & 42.7 & 37.4 & -0.13 & 6.9 & 1 & -2.3 & -7.6\\
1242+1315 & NGC 4639 & 16.8 & 10.97 & 40.0 & 37.1 & 2.29 & 6.6 & 3 & -4.7 & -7.6\\
1313+3635 & NGC 5033 & 18.7 & 14.53 & 41.4 & 38.0 & 1.72 & 7.5 & 3 & -4.2 & -7.6\\
1338+0432 & NGC 5252 & 98.9 & 14.38 & 41.3 & 39.0 & 2.83 & 8.0 & 2 & -4.8 & -7.1\\
1342+3539 & NGC 5273 & 21.3 & 13.51 & 41.0 & 36.8 & 0.95 & 6.5 & 2,3 & -3.6 & -7.8\\
1349-3018 & IC 4329A & 70.2 & 19.4 & 43.3 & 39.0 & 0.81 & 6.7 & 1 & -1.5 & -5.8\\
1353+6918 & Mrk 279 & 135.6 & 20.7 & 43.9 & 38.9 & 0.18 & 7.6 & 1 & -1.9 & -6.8\\
1417+2508 & NGC 5548 & 75.2 & 17.44 & 42.6 & 38.7 & 1.25 & 8.1 & 1 & -3.6 & -7.5\\
1436+5847 & Mrk 817 & 140.4 & 18.96 & 43.2 & 38.8 & 0.76 & 7.6 & 1 & -2.6 & -7.0\\
1504+1026 & Mrk 841 & 156.0 & 18.19 & 42.9 & 39.0 & 1.26 & 8.1 & 2 & -3.4 & -7.2\\
1531+0727 & NGC 5940 & 145.2 & 18.27 & 42.9 & 38.7 & 0.99 & 7.7 & 2,3 & -2.9 & -7.1\\
1616+3542 & NGC 6104 & 119.9 & 16.32 & 42.1 & 38.4 & 1.41 & 7.6 & 2 & -3.6 & -7.3\\
2044-1043 & Mrk 509 & 154.1 & 22.63 & 44.6 & 39.2 & -0.34 & 7.8 & 1 & -1.2 & -6.7\\
2303+0852 & NGC 7469 & 71.4 & 17.93 & 42.8 & 39.3 & 1.72 & 6.8 & 1 & -2.2 & -5.6\\
2318+0014 & Mrk 530 & 124.2 & 16.42 & 42.1 & 39.0 & 2.02 & 8.1 & 3 & -4.1 & -7.2\\
\enddata
\tablecomments{
References in column 9: [1] Ho (2002), [2] Woo \& Urry (2002), [3] Chiaberge et al. (2005), [4] Merloni et al. (2003).}
\end{deluxetable}

\clearpage

\begin{deluxetable}{llllllllllll}
\tabletypesize{\scriptsize}
%\rotate
\tablecaption{FR I Radio Galaxies}
\tablewidth{0pt}
\tablehead{
\colhead{IAU} & \colhead{name} & \colhead{$z$} & \colhead{$\log L_B$} & \colhead{$F_5$} & \colhead{ref.} & \colhead{$\log L_R$} & \colhead{$\log {\cal R}$} & \colhead{$\log {\cal M}_{\rm BH}$} & \colhead{ref.} & \colhead{$\log L_B$} & \colhead{$\log L_R$}\\ J2000.0 & & & [erg/s] & [Jy] & to col. 5 & [erg/s] & & [${\cal M}_{\odot}$] & to col. 9 & [$L_{Edd}$] & [$L_{Edd}$]\\ (1) & (2) & (3) & (4) & (5) & (6) & (7) & (8) & (9) & (10) & (11) & (12)}
\startdata
0055+2624 & 3C 28 & 0.1952 & $<$41.4 & 0.15 & 1 & 41.9 & $>$5.63 & 8.1 & 1 & $<$-4.8 & -4.3\\
0057-0123 & 3C 29 & 0.0448 & 41.3 & 2.01 & 1 & 41.7 & 5.51 & 9.1 & 1 & -5.9 & -5.5\\
0057+3021 & NGC 315 & 0.0167 & 41.0 & 0.914 & 2 & 40.5 & 4.55 & 8.9 & 2 & -6.0 & -6.6\\
0107+32224 & 3C 31 & 0.0169 & 40.9 & 1.12 & 2 & 40.6 & 4.76 & 8.6 & 1 & -5.8 & -6.2\\
0123+3315 & NGC 507 & 0.0164 & $<$39.5 & 0.055 & 3 & 39.2 & $>$4.86 & 9 & 2 & $<$-7.6 & -7.9\\
0125-0120 & 3C 40 & 0.018 & $<$40.5 & 1.78 & 1 & 40.8 & $>$5.46 & 7.9 & 2 & $<$-5.5 & -5.2\\
0156+0537 & NGC 741 & 0.0185 & $<$40.2 & 0.28 & 1 & 40.0 & $>$4.96 & 8.7 & 2 & $<$-6.6 & -6.8\\
0223+4259 & 3C 66B & 0.0215 & 41.5 & 1.77 & 2 & 41.0 & 4.63 & 8.5 & 1 & -5.1 & -5.6\\
0308+0406 & 3C 78 & 0.0288 & 42.5 & 3.45 & 1 & 41.5 & 4.13 & 8.9 & 1 & -4.5 & -5.5\\
0318+4151 & 3C 83.1 & 0.0251 & 40.2 & 3.034 & 4 & 41.3 & 6.30 & 8.8 & 1 & -6.7 & -5.6\\
0319+4130 & 3C 84 & 0.0176 & 42.9 & 42.37 & 2 & 42.2 & 4.42 & 9.1 & 1 & -4.3 & -5.0\\
0334-0110 & 3C 89 & 0.1386 & $<$40.9 & 0.72 & 1 & 42.3 & $>$6.52 & 8.6 & 1 & $<$-5.8 & -4.5\\
1145+1936 & 3C 264 & 0.0206 & 42.0 & 2.36 & 1 & 41.1 & 4.18 & 8.3 & 1 & -4.4 & -5.4\\
1219+0549 & 3C 270 & 0.0074 & 39.4 & 4.86 & 1 & 40.5 & 6.20 & 8.6 & 1 & -7.3 & -6.2\\
1225+1253 & 3C 272.1 & 0.0037 & 40.0 & 2.72 & 1 & 39.6 & 4.73 & 8.2 & 1 & -6.3 & -6.7\\
1230+1223 & 3C 274 & 0.0037 & 41.0 & 67.6 & 1 & 41.0 & 5.12 & 9.5 & 1 & -6.6 & -6.6\\
1259+2757 & NGC 4874 & 0.0239 & $<$39.3 & 0.084 & 2 & 39.7 & $>$5.52 & 8.6 & 2 & $<$-7.4 & -7.0\\
1338+3851 & 3C 288 & 0.246 & 42.0 & 1.008 & 2 & 42.9 & 6.11 & 8.9 & 1 & -5.0 & -4.1\\
1416+1048 & 3C 296 & 0.0237 & 40.5 & 1.202 & 2 & 40.9 & 5.52 & 8.8 & 1 & -6.4 & -6.0\\
1504+2600 & 3C 310 & 0.054 & 41.3 & 1.26 & 1 & 41.6 & 5.52 & 8 & 1 & -4.9 & -4.5\\
1510+7045 & 3C 314.1 & 0.1197 & $<$41.4 & 0.337 & 2 & 41.8 & $>$5.51 & 7.8 & 1 & $<$-4.5 & -4.1\\
1516+0701 & 3C 317 & 0.0342 & 41.4 & 0.93 & 1 & 41.1 & 4.88 & 9.5 & 1 & -6.3 & -6.5\\
1628+3933 & 3C 338 & 0.0303 & 41.2 & 0.477 & 2 & 40.7 & 4.65 & 9.1 & 1 & -6.0 & -6.5\\
1643+1715 & 3C 346 & 0.162 & 43.1 & 1.39 & 2 & 42.7 & 4.74 & 8.8 & 1 & -3.8 & -4.2\\
1651+0459 & 3C 348 & 0.154 & 41.6 & 9.529 & 2 & 43.5 & 7.03 & 8.9 & 1 & -5.4 & -3.5\\
2048+0701 & 4C 424 & 0.127 & $<$41.7 & 0.785 & 2 & 42.2 & $>$5.68 & 8.3 & 1 & $<$-4.7 & -4.2\\
2155+3800 & 3C 438 & 0.29 & $<$41.9 & 1.703 & 2 & 43.3 & $>$6.58 & 8.6 & 1 & $<$-4.8 & -3.4\\
2214+1350 & 3C 442 & 0.0262 & 40.0 & 0.76 & 1 & 40.8 & 5.89 & 8 & 1 & -6.1 & -5.3\\
2231+3921 & 3C 449 & 0.0181 & 41.0 & 0.566 & 2 & 40.3 & 4.47 & 8 & 1 & -5.1 & -5.8\\
2320+0813 & NGC 7626 & 0.0113 & $<$39.8 & 0.21 & 1 & 39.5 & $>$4.85 & 9 & 2 & $<$-7.4 & -7.6\\
2338+2701 & 3C 465 & 0.0301 & 41.5 & 2.12 & 1 & 41.3 & 5.02 & 8.6 & 1 & -5.3 & -5.4\\
\enddata
\tablecomments{
References in column 6: [1] Wright \& Otrupcek (1990), [2] Gregory \& Condon (1991), [3] White \& Becker (1992), [4] Kuhr et al. (1981). References in column 10: [1] Cao \& Rawlings (2004), [2] Woo \& Urry (2002).}
\end{deluxetable}

\clearpage

\begin{deluxetable}{lllllllllll}
\tabletypesize{\scriptsize}
%\rotate
\tablecaption{PG Quasars}
\tablewidth{0pt}
\tablehead{
\colhead{IAU} & \colhead{name} & \colhead{$z$} & \colhead{$\log L_B$} & \colhead{$F_5$} & \colhead{$\log L_R$} & \colhead{$\log {\cal R}$} & \colhead{$\log {\cal M}_{\rm BH}$} & \colhead{ref.} & \colhead{$\log L_B$} & \colhead{$\log L_R$}\\ J2000.0 & & & [erg/s] & [Jy] & [erg/s] & & [${\cal M}_{\odot}$] & to col. 8 & [$L_{Edd}$] & [$L_{Edd}$]\\
(1) & (2) & (3) & (4) & (5) & (6) & (7) & (8) & (9) & (10) & (11)}
\startdata
0010+1058 & PG 0007+106 & 0.089 & 44.3 & 0.321 & 41.5 & 2.29 & 8.3 & 1 & -2.1 & -4.9\\
0029+1316 & PG 0026+129$^{\star}$ & 0.142 & 45.2 & 0.0051 & 40.1 & 0.03 & 7.6 & 2 & -0.5 & -5.6\\
0053+1241 & PG 0050+124$^{\star}$ & 0.061 & 44.7 & 0.0026 & 39.1 & -0.48 & 7.3 & 1 & -0.7 & -6.3\\
0054+2525 & PG 0052+251 & 0.155 & 45.1 & 7.4E-4 & 39.4 & -0.62 & 8.4 & 2 & -1.4 & -7.1\\
0159+0023 & PG 0157+001 & 0.164 & 45.2 & 0.008 & 40.5 & 0.33 & 8.0 & 1 & -0.9 & -5.7\\
0810+7602 & PG 0804+761 & 0.1 & 44.8 & 0.00238 & 39.5 & -0.22 & 8.2 & 2 & -1.5 & -6.9\\
0847+3445 & PG 0844+349 & 0.064 & 44.9 & 3.1E-4 & 38.2 & -1.56 & 7.4 & 2 & -0.6 & -7.3\\
0925+1954 & PG 0923+201 & 0.19 & 45.0 & 2.5E-4 & 39.1 & -0.84 & 8.9 & 2 & -2.0 & -8.0\\
0926+1244 & PG 0923+129$^{\star}$ & 0.029 & 43.8 & 0.01 & 39.0 & 0.32 & 7.0 & 1 & -1.3 & -6.1\\
0956+4115 & PG 0953+414 & 0.239 & 45.7 & 0.0019 & 40.2 & -0.36 & 8.2 & 2 & -0.7 & -6.2\\
1014+0033 & PG 1012+008 & 0.185 & 45.1 & 1E-3 & 39.7 & -0.30 & 8.1 & 1 & -1.1 & -6.6\\
1051-0051 & PG 1049-005 & 0.357 & 45.7 & 4.8E-4 & 40.0 & -0.60 & 9.1 & 1 & -1.6 & -7.3\\
1104+7658 & PG 1100+772 & 0.313 & 45.6 & 0.66 & 43.0 & 2.51 & 9.1 & 1 & -1.6 & -4.2\\
1106-0052 & PG 1103-006 & 0.425 & 45.8 & 0.482 & 43.2 & 2.44 & 9.3 & 1 & -1.6 & -4.2\\
1117+4413 & PG 1114+445 & 0.144 & 44.8 & 2.2E-4 & 38.8 & -0.89 & 8.4 & 1 & -1.7 & -7.7\\
1119+2119 & PG 1116+215 & 0.177 & 45.3 & 0.0028 & 40.1 & -0.14 & 8.5 & 1 & -1.3 & -6.5\\
1121+1144 & PG 1119+120$^{\star}$ & 0.049 & 44.4 & 9.4E-4 & 38.4 & -0.82 & 7.2 & 1 & -0.9 & -6.9\\
1204+2754 & PG 1202+28 & 0.165 & 45.3 & 8.3E-4 & 39.5 & -0.73 & 8.1 & 1 & -0.9 & -6.8\\
1214+1403 & PG 1211+143$^{\star}$ & 0.085 & 44.9 & 8E-4 & 38.9 & -0.9 & 7.5 & 2 & -0.7 & -6.8\\
1219+0638 & PG 1216+069 & 0.334 & 45.7 & 0.004 & 40.8 & 0.22 & 9.2 & 1 & -1.6 & -6.4\\
1232+2009 & PG 1229+204 & 0.064 & 44.6 & 6.7E-4 & 38.5 & -0.97 & 8.6 & 2 & -2.1 & -8.2\\
1246+0222 & PG 1244+026$^{\star}$ & 0.048 & 43.8 & 8.3E-4 & 38.4 & -0.28 & 6.3 & 1 & -0.6 & -6.1\\
1301+5902 & PG 1259+593 & 0.472 & 46.0 & 3E-5 & 39.1 & -1.86 & 9.0 & 1 & -1.1 & -8.0\\
1309+0819 & PG 1307+085 & 0.155 & 45.2 & 3.5E-4 & 39.0 & -1.00 & 7.9 & 2 & -0.8 & -7.0\\
1312+3515 & PG 1309+355 & 0.184 & 45.3 & 0.054 & 41.4 & 1.26 & 8.2 & 1 & -1.1 & -4.9\\
1353+6345 & PG 1351+640 & 0.087 & 44.6 & 0.0133 & 40.1 & 0.64 & 8.5 & 2 & -2.0 & -6.5\\
1354+1805 & PG 1352+183 & 0.158 & 45.0 & 2.5E-4 & 38.9 & -0.97 & 8.3 & 1 & -1.4 & -7.5\\
1405+2555 & PG 1402+261$^{\star}$ & 0.164 & 45.1 & 6.2E-4 & 39.4 & -0.64 & 7.3 & 2 & -0.3 & -6.1\\
1413+4400 & PG 1411+442 & 0.089 & 44.8 & 6.1E-4 & 38.8 & -0.87 & 7.6 & 2 & -0.9 & -6.9\\
1417+4456 & PG 1415+451 & 0.114 & 44.7 & 4E-4 & 38.8 & -0.76 & 7.8 & 1 & -1.2 & -7.1\\
1419-1310 & PG 1416-129 & 0.129 & 44.9 & 0.0036 & 39.9 & 0.06 & 8.5 & 1 & -1.7 & -6.7\\
1429+0117 & PG 1426+015 & 0.086 & 44.7 & 0.00121 & 39.0 & -0.55 & 7.9 & 2 & -1.3 & -7.0\\
1442+3526 & PG 1440+356$^{\star}$ & 0.077 & 44.6 & 0.00166 & 39.1 & -0.44 & 7.3 & 1 & -0.8 & -6.3\\
1446+4035 & PG 1444+407 & 0.267 & 45.4 & 1.6E-4 & 39.2 & -1.07 & 8.2 & 1 & -0.9 & -7.1\\
1535+5754 & PG 1534+580 & 0.03 & 43.6 & 0.00192 & 38.3 & -0.16 & 7.4 & 1 & -1.9 & -7.2\\
1547+2052 & PG 1545+210 & 0.266 & 45.4 & 0.72 & 42.9 & 2.62 & 9.1 & 1 & -1.9 & -4.4\\
1613+6543 & PG 1613+658 & 0.129 & 44.9 & 0.00303 & 39.8 & 0 & 8.6 & 2 & -1.8 & -6.9\\
1620+1724 & PG 1617+175 & 0.114 & 44.8 & 0.00109 & 39.3 & -0.38 & 7.9 & 2 & -1.2 & -6.7\\
1701+5149 & PG 1700+518 & 0.292 & 45.7 & 0.0072 & 41.0 & 0.37 & 8.3 & 2 & -0.7 & -5.5\\
2132+1008 & PG 2130+099 & 0.061 & 44.6 & 0.00205 & 39.0 & -0.50 & 7.7 & 2 & -1.3 & -6.9\\
2211+1841 & PG 2209+184 & 0.07 & 44.2 & 0.29 & 41.2 & 2.15 & 8.5 & 1 & -2.4 & -5.4\\
2254+1136 & PG 2251+113 & 0.323 & 45.5 & 0.523 & 42.9 & 2.56 & 9.0 & 1 & -1.6 & -4.1\\
2311+1008 & PG 2308+098 & 0.432 & 45.8 & 0.303 & 43.0 & 2.27 & 9.6 & 1 & -1.9 & -4.7\\
\enddata
\tablecomments{
References in column 9: [1] Vestergaard (2002), [2] Woo \& Urry (2002). Symbol `$^{\star}$'
in column 2 denotes objects classified as Narrow-Line Seyfer 1 Galaxies. }
\end{deluxetable}

\clearpage

\begin{figure}
\centering
\includegraphics[scale=1.3]{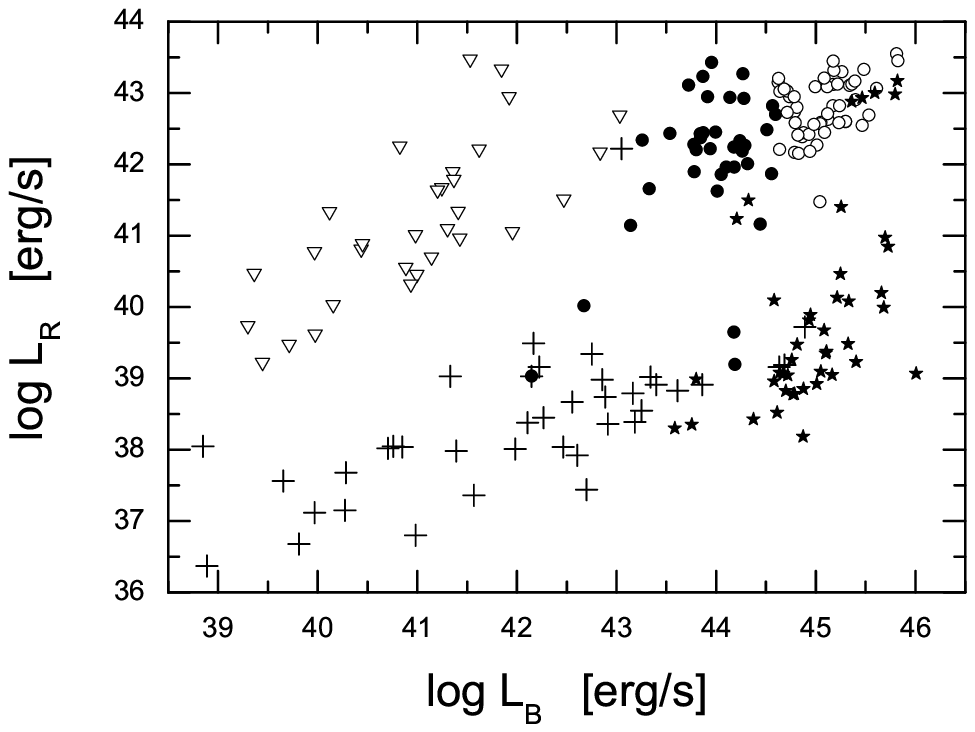}
\caption{Total $5$ GHz luminosity vs. $B$-band nuclear luminosity.
BLRGs are marked by filled circles, radio-loud quasars by open
circles, Seyfert galaxies and LINERs by crosses, FR I radio galaxies
by open triangles, and PG Quasars by filled stars.}
\label{fig1}
\end{figure}

\clearpage

\begin{figure}
\centering
\includegraphics[scale=1.3]{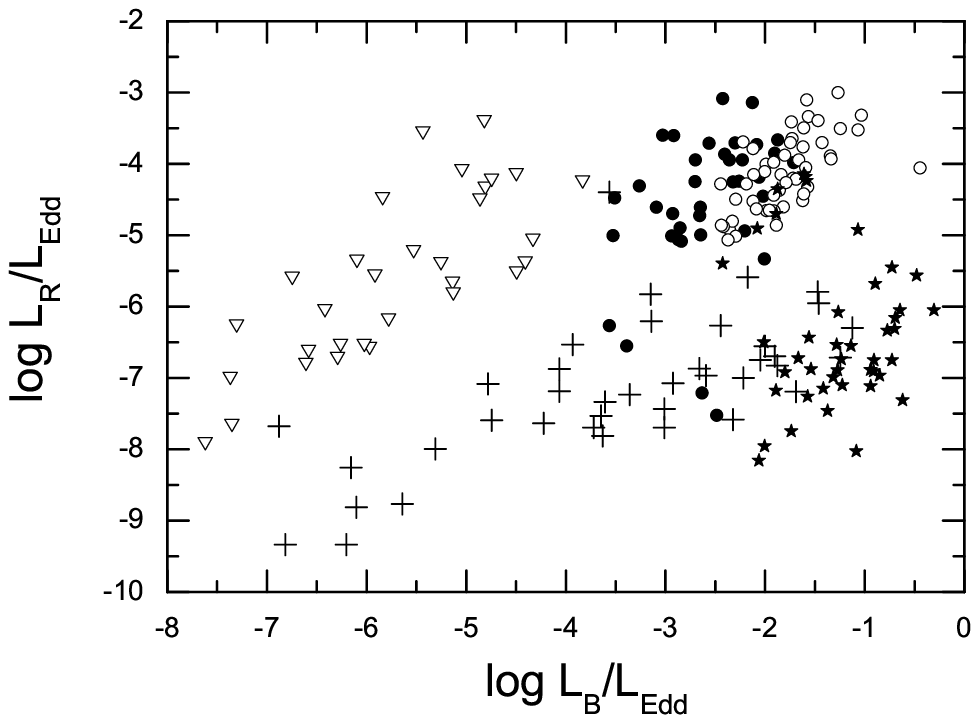}
\caption{Total $5$ GHz luminosity vs. $B$-band nuclear luminosity
in the Eddington units. BLRGs are marked by the filled circles,
radio-loud quasars by the open circles, Seyfert galaxies and LINERs
by the crosses, FR I radio alaxies by the open triangles, and PG
Quasars by the filled stars.} \label{fig2}
\end{figure}

\clearpage

\begin{figure}
\centering
\includegraphics[scale=1.3]{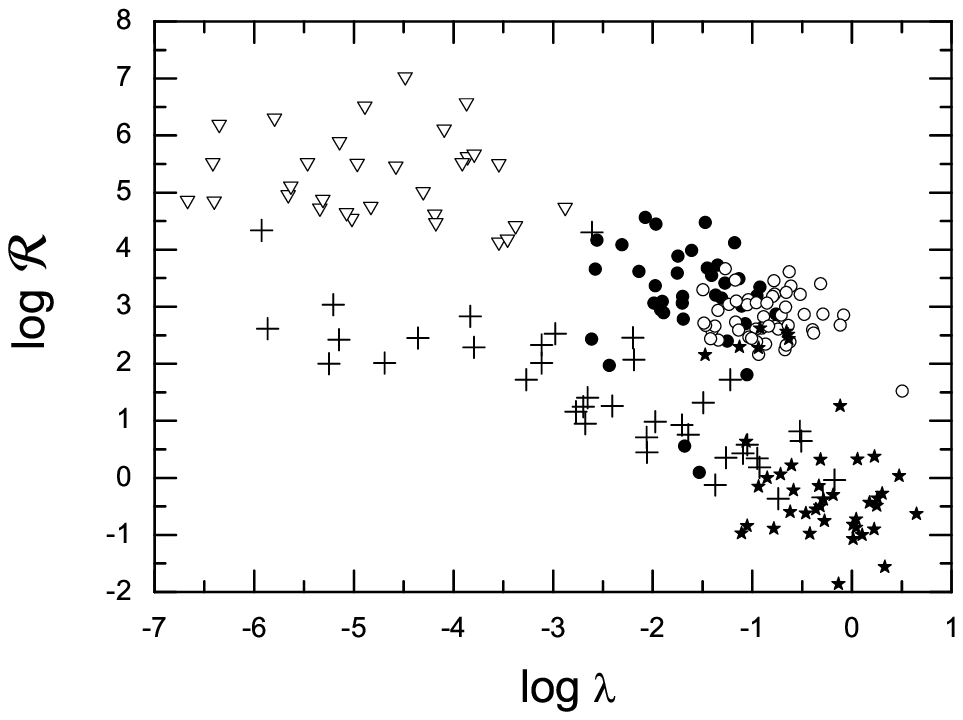}
\caption{Radio-loudness ${\cal R}$ vs. Eddington ratio $\lambda$.
BLRGs are marked by the filled circles, radio-loud quasars by the
open circles, Seyfert galaxies and LINERs by the crosses, FR I radio
alaxies by the open triangles, and PG Quasars by the filled stars.}
\label{fig3}
\end{figure}

\clearpage

\begin{figure}
\centering
\includegraphics[scale=1.3]{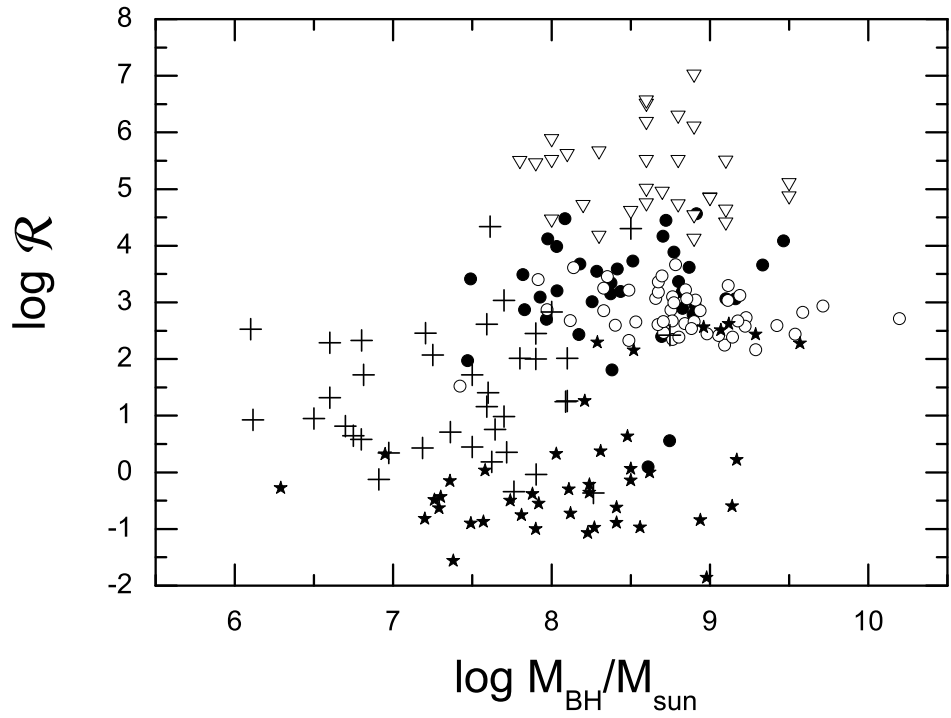}
\caption{Radio-loudness vs. black hole mass. BLRGs are marked by the
filled circles, radio-loud quasars by the open circles, Seyfert
galaxies and LINERs by the crosses, FR I radio galaxies by the open
triangles, and PG Quasars by the filled stars.} \label{fig4}
\end{figure}

\end{document}